\documentclass[aps,rmp,reprint,amsmath,amssymb,graphicx,longbibliography]{revtex4-1}

\usepackage{amsmath}
\usepackage{amssymb}
\usepackage{graphicx}
\usepackage{dcolumn}
\usepackage{bm}
\usepackage{bbold}
\usepackage{xcolor}
\usepackage{soul}
\usepackage{caption}
\usepackage[english]{babel}
\usepackage{listings}
\DeclareMathOperator{\argmax}{arg\,max}

\begin{document}
\captionsetup[figure]{font=small,labelformat={default},labelsep=period,name={FIG.}}

\title{Multiscale modeling of brain network organization}

\author{Charley, Presigny}
\affiliation{Inria Paris, Aramis Project Team, Paris, France\\}
\affiliation{Sorbonne University, UPMC Univ Paris 06, Inserm U-1127, CNRS UMR-7225, Paris Brain Institute (ICM), Hopital Pitie-Salpetriere, Paris, France\\}

\author{Fabrizio, De Vico Fallani}%
\email{fabrizio.de-vico-fallani@inria.fr}
\affiliation{Inria Paris, Aramis Project Team, Paris, France\\}
\affiliation{Sorbonne University, UPMC Univ Paris 06, Inserm U-1127, CNRS UMR-7225, Paris Brain Institute (ICM), Hopital Pitie-Salpetriere, Paris F-75013, France\\}

\date{\today}

\begin{abstract}
A complete understanding of the brain requires an integrated description of the numerous scales and levels of neural organization. That means studying the interplay of genes and synapses, but also the relation between the structure and dynamics of the whole brain, which ultimately leads to different types of behavior, from perception to action, while asleep or awake.
Yet, multiscale brain modeling is challenging, in part because of the difficulty to simultaneously access information from multiple scales and levels. While some insights have been gained on the role of specific microcircuits on the generation of macroscale brain activity, a comprehensive characterization of how changes occurring at one scale/level, can have an impact on other ones, remains poorly understood.
Recent efforts to address this gap include the development of new frameworks mostly originating from network science and complex systems theory. These theoretical contributions provide a powerful framework to analyze and model interconnected systems exhibiting interactions within and between different \textit{layers} of information. Here, we present recent advances for the characterization of the multiscale brain organization in terms of structure-function, oscillation frequencies and temporal evolution. Efforts are reviewed on the multilayer network properties underlying the physics of higher-order organization of neuronal assemblies, as well as on the identification of multimodal network-based biomarkers of brain pathologies, such as Alzheimer's disease. 
We conclude this Colloquium with a perspective discussion on how recent results from multilayer network theory, involving generative modeling, controllability and machine learning, could be adopted to address new questions in modern physics and neuroscience.

\end{abstract}

\maketitle

\tableofcontents

\newpage

\section{\label{sec:intro}Introduction}

\begin{figure*}
\includegraphics[scale=0.5]{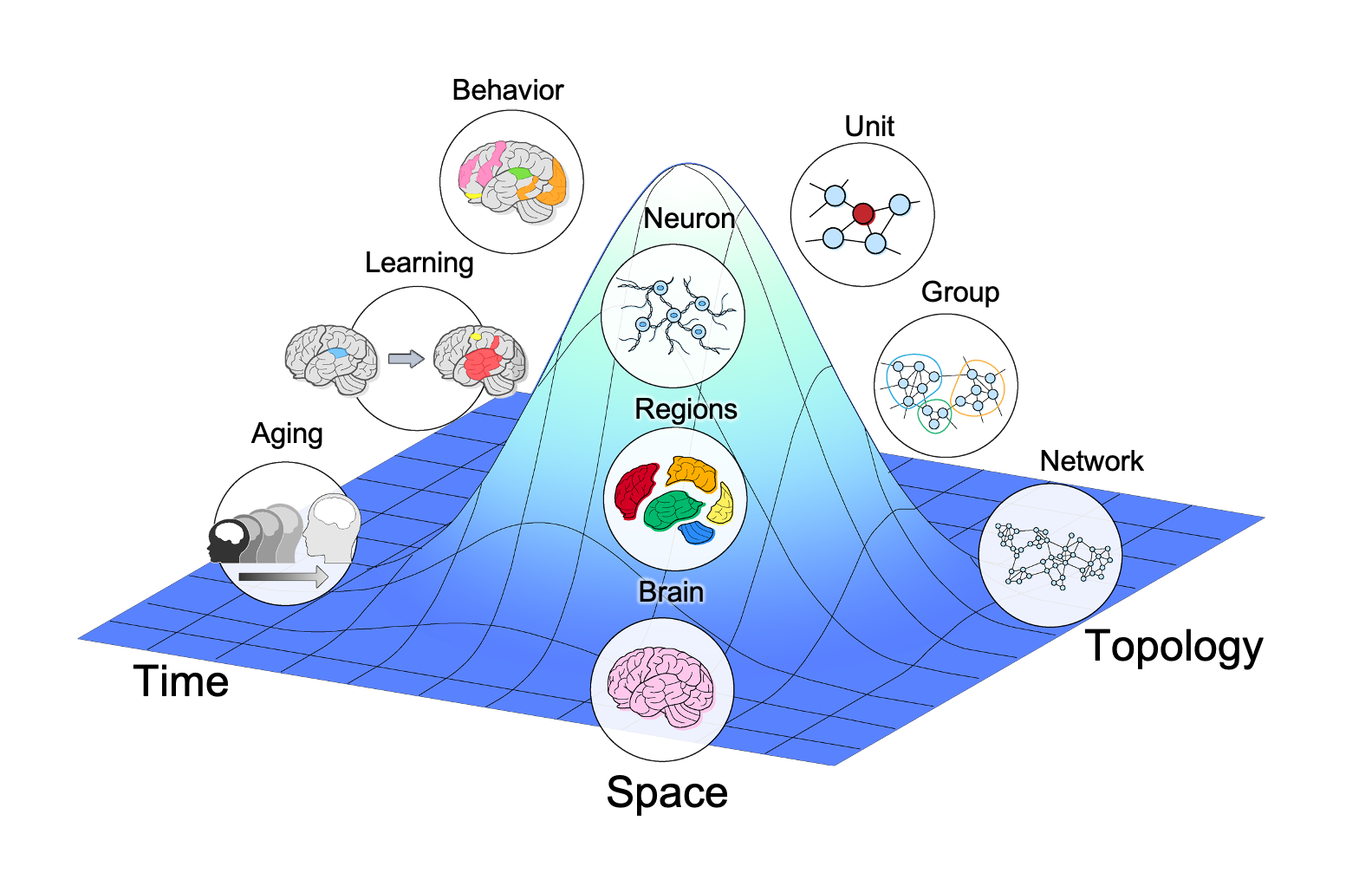}
\caption{\label{fig:multiscale} \textbf{Multiscale brain organization.} The different organizational aspects of the brain system are represented over a multidimensional manifold. Three type of dimensions, 
or levels, are illustrated here, i.e. time, space, and topology. From the top to the bottom of the manifold, the scales of each organizational level go from micro to macro. Image credit: Thibault Rolland, t.d.rolland@gmail.com}
\end{figure*}

The brain is a formidable complex system exhibiting a wide repertoire of emergent phenomena, such as criticality, that ultimately rule the behavior of many living beings \citep{beggs2003,fontenele2019,chialvo2010,arcangelis_2006,WILTING2019105}. 
Those phenomena involve multiple spatial scales, from molecules to the whole brain, and stem from multiple temporal scales, from sub-milliseconds to the entire lifespan \citep{robinson_multiscale_2005}. 
More broadly, scales can refer to other types of dimensions or levels \citep{levels_2021}, such as complementary phenomenological information captured by different experimental technologies (e.g., magnetic resonance imaging, electrophysiology, genetics) or neuronal interactions at multiple topological levels \citep{betzel2017,bazinet2021}
(\textbf{Fig.~\ref{fig:multiscale}}).

Disentangling such organizational complexity and investigating how the relationships between the system's parts give rise to its collective behavior, is crucial to understand basic neural functioning \citep{turkheimer_2021} and, eventually, cure brain diseases \citep{cutsuridis2019}. 
Modeling multiscale brain organization is indeed one of the most important challenges of our century. 
The number of flagship initiatives funding large projects that aim to reproduce multiscale brain behavior has significantly increased over the last two decades, i.e., Human Brain Project\footnote{humanbrainproject.eu}, the Brain Initiative\footnote{braininitiative.nih.gov}, and China Brain Project \citep{china} just to cite few examples. 

While at present there is no comprehensive theory of how to bridge multiple scales and levels, the pursuit of such a theory remains critically important. Several recent models propose new ways to model neural activity within and between multiple scales, and further provide mechanistic insights into the structure and dynamics of brain organization. Hence, it is timely to discuss these emerging developments, and to seek to tie them together into a meaningful theoretical field that tackles current open questions in multiscale neuroscience and medicine from a system perspective.

Research in the field has progressively acknowledged the importance of considering brain organization from a holistic perspective and not from a reductionist angle \citep{breakspear_dynamic_2017,deco2011,engel_2021}. 
This is somewhat implicit in the term organization itself, which stems from the Medieval Latin \textit{organizatio} i.e., the arrangement of parts in an organic whole.
Accumulating evidence indicates that modeling how different brain components interact is often more realistic and effective in terms of behavior prediction, than simply considering their activity in isolation  \citep{scannell_connectional_1999,friston_2011}.

Graphs (or networks) have progressively emerged as a natural way to describe heterogeneous connectivity diagrams at single scales or levels \citep{jouve1998,sporns2000,hilgetag_clustered_2004,stam2007,friston2013}.
According to this framework, the nodes of a network correspond to different brain sites, such as neurons, neuronal ensembles or even larger areas but also to electric or optical sensors.
The edges, or links, of the network represent either anatomical/structural connections or functional/dynamical interactions between the nodes.
While the best practices for establishing the links between brain nodes are still evolving, the type of connectivity basically depends on the experimental technology. 
Anatomical brain networks are often derived from post-mortem tract tracing or \textit{in-vivo/vitro} structural imaging (e.g., diffusion tensor imaging DTI) \citep{rubinov2010}.
Dynamical brain networks are instead mostly obtained from \textit{in-vivo/vitro} functional imaging, such as optical imaging, electrophysiology (e.g., EEG, MEG), or functional magnetic resonance imaging (fMRI) \citep{fabrizio2014}.

The use of a network formalism to study the structure and dynamics of interconnected brain systems has a rich and pervasive heritage in seminal works at the intersection between physics and neuroscience. 
Studies on single-scale brain networks brought up major results and got structured around concepts and languages inspired by statistical physics and complex systems theory.
Similarly to other real interconnected systems, brain networks tend to exhibit an optimal balance between integration and segregation within their connectivity structure \citep{bassett_revisited_2017}. 
This peculiar structure, also known as \textit{small-world}, is topologically characterized by the co-occurence of short paths and abundant clustering links between nodes \citep{watts1998}. Small-world networks ensure efficient communication between the nodes and favor global synchronization of oscillatory dynamics \citep{latora2001,oscillation2000}.

Brain networks also exhibit other important topological properties, such as mesoscale modular organization as well as the presence of core hubs passing information between peripheral distant brain areas \citep{bullmore2009,kennedy2013,heuvel2011,zamora2010}. 
In addition, being embedded in space, brain networks are economic as they tend to minimize the energetic cost (e.g., metabolic) associated to the presence of long-range connections \citep{bullmore_economy_2012}. 

At this stage, it is important to remind that the brain is a flexible system, and its organization can adapt to the external environment, endo/exogenous inputs, as well as during diseases or after damages. As a consequence, topological properties of brain networks can exhibit shifts from normative physiological values and those deviations constitute the basis for the identification of new organizational mechanisms and biomarkers in both cognitive and clinical neuroscience \citep{medaglia_cognitive_2015,stam_modern_2014,zalesky2014,fornito_connectomics_2015,fornito_genetic_2021}. 

All the aforementioned findings refer to brain networks obtained separately from different levels of information. Here, we expand the link between physics and neuroscience by building a unifying framework to analyze and model neural organization across multiple scales and/or dimensions from a network perspective.
Specifically, we will focus on approaches based on multilayer network theory, a recent field connected to physics through nontrivial results related to structure, dynamics, criticality and resilience \citep{dedomenico2013,boccaletti2014,zanin2015,radicchi_2014,della_rossa_symmetries_2020,aleta_2019, nicozia_2017, danziger_recovery_2022}.

In addition, network models based on hyperbolic geometry have been recently introduced to reproduce brain networks at different coarse-grained spatial resolutions, providing new insights on brain self-similarity and criticality \citep{allard_hyperbolic,zheng_renormalization}.
More traditionally, multiscale brain modeling can be performed by designing biophysical models of single scale dynamics and simulating simple inter-layer connectivity schemes \citep{siettos2016,lytton_multiscale_2017,cutsuridis2019}.
Should the readers be interested in these alternative approaches we direct them to the above mentioned references.

\begin{figure*}
\includegraphics[scale=0.4]{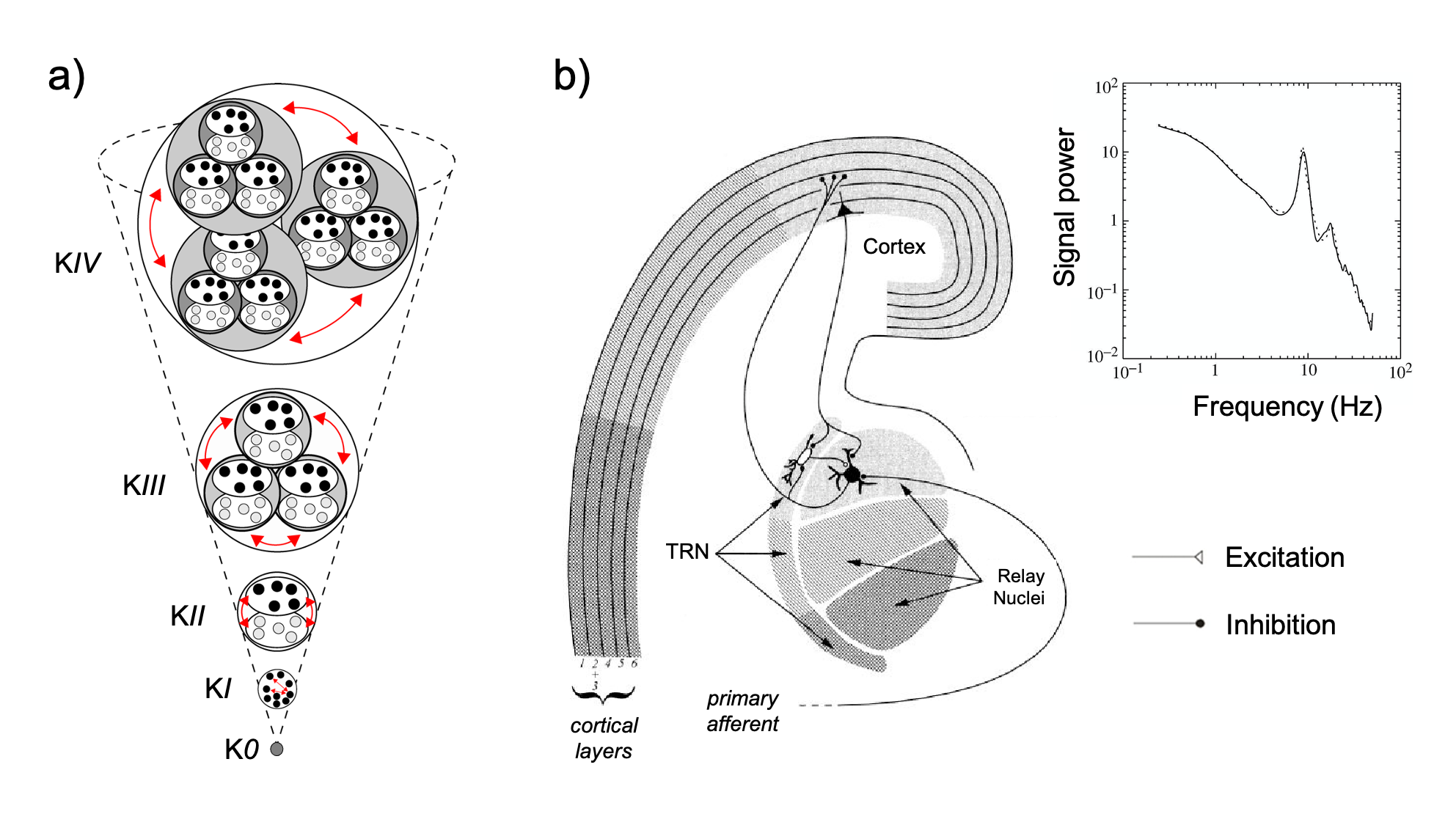}
\caption{\label{fig:bottom_up} \textbf{Bottom-up hierarchical modeling.}
Panel a) The so-called K-set hierarchy showing the model progression from cell level to entire brain. K0 is a noninteracting collection of neurons. KI corresponds to a cortical column with sufficient functional connection density. KII represents a collection of excitatory and inhibitory populations. KIII is formed by the interaction of several KII sets and simulates the known dynamics of sensory areas with 1/f spectra (see inset b). KIV is formed by the interaction of three KIII sets that models the genesis of simple forms of intentional behaviors. 
Panel b) Schematic view of major components involved in thalamocortical interactions. Different shading patterns code for different zones of the system, i.e. from micro (relay nuclei, thalamic reticular nuclei (TRN)) to macro scales (cortex). As indicated by key, all connections shown are excitatory except for the connection from the reticular cell to the relay cell, which is inhibitory.
Pictures and captions adapted from \citet{kozma2007} (panel a) and \citet{sherman1996} (panel b). Inset b republished from \citet{robinson_multiscale_2005} with permission of The Royal Society (U.K.); permission conveyed through Copyright Clereance Center, Inc.}
\end{figure*}

The remainder of this Colloquium is organized as follows. 
In Sec.~II we illustrate the rationale of multiscale brain modeling, and briefly review the main research lines and challenges. These arguments anticipate the introduction of the multilayer network theory to characterize brain network organization across multiple scales and levels.
In Sec.~III, we present the multilayer network formalism by providing basic notions and definitions. We then introduce ways of characterizing multilayer network properties that have been extended to network neuroscience. Sec.~IV describes the different types of multilayer brain networks that have been investigated so far. Emphasis is given on the relevance of multilayer modeling as compared to single-layer alternatives and on the current practices to infer them from experimental data.
We next turn into Sec.~V, to detail a few examples of how multilayer network theory has been used to characterize and understand brain structure and function in physiological conditions.
Then, in Sec.~VI, we describe which multilayer network properties deviate from normative values in the presence of brain diseases, and how to derive predictive biomarkers of network reorganization associated with clinical outcomes. We close in Sec.~VII by outlining the emerging frontiers of multilayer network theory that can be developed to advance multiscale brain modeling. 
Except otherwise stated, brain networks refer here to connectivity graphs obtained with neuroimaging techniques in humans.
Nonetheless, the presented formalism is broadly relevant and applicable to other animal species (primates and non-primates), data modalities (\textit{in-vitro/vivo}), as well as to simulated neural models (\textit{in-silico}).

By reviewing the research endeavors on multilayer network theory to study the brain, we aim to stimulate a discussion and reflection on the exciting opportunity it constitutes for multiscale neural modeling. To this end, we kept to the minimum jargon terminology and we adopted an accessible language to reach the broadest multidisciplinary science community. 

\section{Multiscale brain modeling}

The goal of multiscale modeling is to describe system's behavior by simultaneously considering multiple features, or mechanisms, taking place and interacting on different levels of information.
These levels may represent phenomena of different nature, such as in continuum mechanics and molecular dynamics, or at different spatio/temporal resolution i.e., from micro to macro scales.
Multiscale modeling is therefore central for the integrated understanding of a system complexity and for the prediction of its emergent properties.
Since most real-life phenomena involve a broad range of spatial or temporal scales, as well as the interaction between different processes, multiscale modeling has been widely adopted in several disciplines, from material science and algorithmics, to biology and engineering \citep{weinan2011}.

In neuroscience, multiscale modeling has historically considered multiple levels ranging from microscopic single neuron activity to macroscopic behavior of collective dynamics. This is achieved by bridging biophysical mechanistic models of neuron dynamics and experimental neuroimaging data \citep{gerstner2012}.
Such “bottom-up” approach allows to predict macroscopic observables by integrating information at smaller scales, typically under the assumption of mean-field approximations \citep{breakspear2005,siettos2016,goldman_2019}. This means that neuronal ensembles' dynamics are progressively “averaged" across scales leading to a characteristic hierarchical nested structure where multiple units at finer-grained levels map into a new entity at coarser-grained ones \textbf{Fig.~\ref{fig:bottom_up}a)} \citep{freeman1975,kozma2003,chialvo2010,expert2010}.

The thalamocortical model is perhaps one of the simplest examples that can reproduce disparate physiological and pathological conditions, from Parkinson’s disease to epileptic seizures \citep{bhatta2011,sohanian_haghighi_new_2017,sherman1996,bonjean2012,jirsa1996, lopes_da_silva_model_1974}. 
In this model, both basic microscopic neurophysiology (e.g., synaptic and dendritic dynamics) and mesoscale brain anatomy (e.g., corticocortical and corticothalamic pathways) are progressively incorporated to predict large-scale brain electrical activity (\textbf{Fig.~\ref{fig:bottom_up}b}).

With the advent of new technologies and tools that allow gathering more precise experimental data and efficient processing, multiscale brain modeling has witnessed a significant transformation in the last decade. Increasingly more sophisticated and accurate models have been proposed including, among others, large-scale anatomical and functional brain connectivity \citep{deco2008,deco2011}.
However, to fully understand a multiscale system, models at different scales must be coupled together to produce integrated models across multiple levels. Indeed, global brain dynamics are strongly dependent on the interaction of several interconnected subnetworks that differently contribute to generate them.
Thus, the study of how intra-scale and inter-scale interactions give rise to collective behavior and to relationships with their environment is a central theme of modern multiscale brain modeling.
Because of the substantial lack of biological evidence, especially concerning inter-scale connectivity, large parts of the studies have focused on analytical and numerical approaches \citep{dada2011}. For example, intra-scale interactions have been simulated adopting cellular automata perspectives \citep{kozma2004}, while inter-scale connectivity have been established using wavelet transformations \citep{breakspear2005}. 

The use of “top-down” approaches, which start with the observation of biological characteristics in the intact system and then construct theories that would explain the observed behaviors, offers complementary solutions. In particular, data-driven methods based on statistical signal/image processing of neuroimaging data allow to infer network representations of the brain at both anatomical and functional levels. 
Structural/anatomical brain networks are typically derived from 3D modeling techniques which identify nerve tracts using data collected by diffusion MRIs \citep{basser_2000}.
Functional/dynamical brain networks are instead estimated by computing similarities between the activity signals generated in different brain sites. To this end, related measures such as Pearson or Spearman correlations, mutual information or Granger-causality are used depending on the nature of the experiment and on the type of scientific question \citep{fabrizio2014}. 
Interestingly, the use of cross-frequency coupling represents a promising approach to derive inter-scale interactions across multiple signal oscillation frequencies \citep{jirsa2013}. 
Thus, while multiscale modeling in neuroscience has historically had a strong spatial connotation, it nowadays spans disparate levels of information, from structure/function to multiple oscillatory regimes and temporal evolution. 
Top-down approaches can be therefore used to generate richer and more realistic models reproducing real brain connectivity schemes and not just simulated ones \citep{siettos2016}.

However, richer information and more accurate models also mean higher complexity and harder interpretation. These are both typical characteristics of multiscale problems that require the use of efficient algorithms to simulate the fully integrated model and appropriate ways of analyzing and interpreting them \citep{chi_neural_2016}. This is actually one of the main challenges of big research projects supported by funding agencies around the world, such as the European Human Brain Project\footnote{humanbrainproject.eu} or the US BRAIN Initiative\footnote{braininitiative.nih.gov}. The increasing number of open-source tools that can be freely accessed and customized to enrich multiscale brain models just confirms how broad and multidisciplinary is the community effort \citep{neuron,nest,netpyne,leon2013}.

In all this turmoil, questions like: how to model within-level and between-level relationships, how to characterize the resulting higher-order network properties, what are the critical phenomena emerging from the interaction of multiple levels, appear to be essential for advancing multiscale models. 
These questions and associated notions motivate the construction of a theory that explicitly builds on the capability to simultaneously characterize intralayer and interlayer connectivity. In the next section, we will introduce the methodological framework of multilayer network theory, which is at the basis of recent developments in multiscale modeling of neural functioning.

\section{Multilayer network formalism}

\subsection{Mathematical definition of multilayer networks}

\begin{figure*}

\includegraphics[scale=0.5]{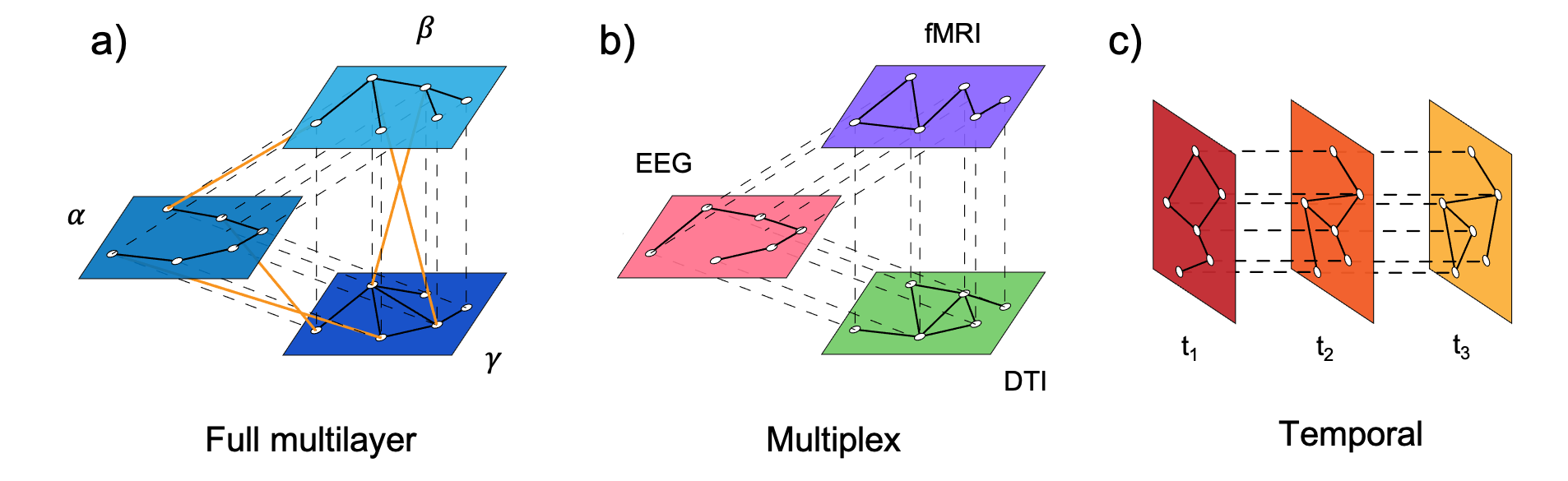}
\caption{\label{fig:type_multilayer} \textbf{Main configurations of multilayer networks.} 
Panel a) Full multilayer network. Both within- and between-layer connections are allowed with no specific restrictions. This configuration is typically adopted to model multifrequency brain networks (see Sec.~\ref{sec:type_multilayer}). 
Panel b) Multiplex network. Only interlayer connections between the replica nodes are allowed. No restrictions on connections within layers. This configuration is typically used to model multimodal brain networks (see Sec.~\ref{sec:type_multilayer}). Panel c) Temporal network. Interlayer connections are allowed only between adjacent layers. No restrictions on connections within layers. This configuration is typically adopted to model time-varying brain networks (see Sec.~\ref{sec:type_multilayer}). Image credit: Thibault Rolland, t.d.rolland@gmail.com}

\end{figure*}



The need to investigate complex systems with multiple types of connectivity has emerged, almost independently, from different disciplines including social science, engineering, and computer science \citep{wasserman1994,little2002,dunlavy2011}. More recently, the physics community also produced pioneering works on various notions such as networks of networks \citep{zhou2006,zhou2007},
node-colored networks \citep{newman2003,vazquez2006}, interdependent networks \citep{buldyrev2010,gao2012}  or multilayer networks \citep{jo2006,kurant2006}.
As a consequence, different terms have been introduced and adopted, thus producing a lack of a consensus set of terminology and mathematical formulation. Only in the last decade, we have eventually witnessed the dawning of general frameworks compatible with tools from complex systems and network science \citep{kivela2014,boccaletti2014}, or based on tensorial formalisms \citep{domenico2017}.

Formally, a \textit{multilayer network} is defined as 
$\mathcal{M}=(\mathcal{G},\mathcal{C})$
where $\mathcal{G}$ is a set of graphs and $\mathcal{C}$ a set of edges connecting the nodes of the different graphs \citep{boccaletti2014}.
More precisely, $\mathcal{G} = \{G_\alpha | \alpha \in \mathbb{N}\}$ with $G_\alpha = (V_\alpha, E_\alpha)$ being a graph at layer $\alpha$. $V_\alpha$ is the set of nodes of $G_\alpha$ and $E_\alpha$ the set of its edges, with $E_\alpha \subseteq V_\alpha \times V_\alpha$.
The set of edges between the nodes of the graphs at different layers $\alpha$ and $\beta$ is denoted by $\mathcal{C} = \{E_{\alpha \beta} \subseteq V_\alpha \times V_\beta | \alpha \neq \beta\}$.

An equivalent, but less formal, convenient representation of a multilayer network is given by the so-called \textit{supra-adjacency matrix}, $A = \{a_{ij}^{\alpha \beta}\}$. Here, the element $a_{ij}^{\alpha \beta}$ represents the link between node \textit{i} in layer $\alpha$ and node \textit{j} in layer $\beta$. 
Hence, given $M$ layers in the graph, $A$ will result in a matrix with $M$ blocks on the main diagonal, accounting for the connections within layers, and $M(M-1)$ off-diagonal blocks describing the links between different layers (Eq.~\ref{adjacency}).

The above definitions are quite general and allow to describe complex systems exhibiting different number of nodes in each layer or scale, directed or undirected interactions, as well as weighted or unweighted connectivity.
Based on state-of-the-art studies \cite{bianconi2018}, we consider here multilayer networks composed of \textit{replica} nodes. That means that all the layers will have the same number of nodes representing the same units of the system across different scales. Note that this is a very particular condition, which however matches the nature of data presented in large part of the studies so far.
In this configuration $V_\alpha = V, \alpha \in \{1,...,M\}$ and only connectivity within and between layers is allowed to change (\textbf{Fig.~\ref{fig:type_multilayer}a}). In the following, we will refer to these general configurations as to \textit{full multilayer} networks.
The supra-adjacency matrix of full multilayer networks has the following form:

\begin{align}
\label{adjacency}
A = \left(
\begin{array}{c|c|c|c}
E_{11} & E_{12} & \dots & E_{1M} \\
\hline 
E_{21} & E_{22} & \dots & E_{2M} \\
\hline
\vdots & \ddots & \vdots &\vdots \\
\hline
E_{M1} & E_{M2} & \dots & E_{MM} 
\end{array}
\right),
\end{align}
where $E_{\alpha \beta}$ contains interlayer links when $\alpha \neq \beta$ and intralayer links when $\alpha = \beta $.

Specific cases of full multilayer networks are the so-called \textit{multiplex} networks. In multiplexes, interlayer connections are not present apart from those between replica nodes (\textbf{Fig.~\ref{fig:type_multilayer}b}). These links inform the model of the existing nodal correspondences across layers.
Hence, in a multiplex $V_\alpha = V,\alpha \in \{1,...,M\}$  and $\mathcal{C} = \{E_{\alpha \beta} \subseteq \{(v,v)| v\in V\} | \alpha \neq \beta\}$. The associated supra-adjacency matrix becomes:
\begin{align}
\label{multiplex_adj}
A = \left(
\begin{array}{c|c|c|c}
E_{11} & I & \dots & I \\
\hline 
I & E_{22} & \dots & I \\
\hline
\vdots & \ddots & \vdots &\vdots \\
\hline
I & I & \dots & E_{MM} 
\end{array}
\right),
\end{align}
where $I$ is the $N \times N$ identity matrix.

Based on the above configurations, many types of multiscale interconnected systems (e.g., spatial, temporal, multimodal) can be represented and investigated. 
For example, \textit{temporal} networks are represented by a particular type of multiplex, where only replica nodes between adjacent layers are interconnected, and the blocks after the first diagonals in Eq.~\ref{multiplex_adj} become zero matrices (\textbf{Fig.~\ref{fig:type_multilayer}c}).
We notice that in general the information contained in multilayer networks can be obtained neither from equivalent aggregated versions (e.g., where links are averaged across layers) nor from standard network metrics and tools \cite{zanin2015}.
For this reason, it is crucial to derive new concepts and methods to quantify the higher-order topological properties emerging from multilayer networks. 
In the next subsection, we introduce some of the metrics and tools that have been developed so far, and that have been adopted in neuroscience. For the sake of simplicity, we will focus on unweighted and undirected multilayer networks.
Should the readers be interested in a more complete picture of multilayer network theory, we refer them to recent reviews by \citet{boccaletti2014} and \citet{bianconi2018}.

\subsection{\label{sub:tool}Analytical tools for multilayer networks}

In the following, we briefly present some of the multilayer methods that have been most frequently used in neuroscience.
We categorize them according to the topological scale that they characterize, i.e. from nodes (micro-scale) to the entire network (macro-scale) passing by groups of nodes (meso-scale).
As a reminder, the entry $a_{ij}^{\alpha \beta}=a_{ji}^{\alpha \beta}$ of the supra-adjacency matrix $A$ describes the interaction of node \textit{i} in layer $\alpha$ to node \textit{j} in layer $\beta$ of a given multilayer network. 
Since $A$ is binary, $a_{ij}^{\alpha \beta}$ has either the value of $1$ (presence of a link) or $0$ (absence of a link). 
Note that $a_{ij}^{\alpha}$ represents intralayer interactions in layer $\alpha$. Finally, we denote the number of layers as $M$ and the number of replica nodes in each layer as $N$.

\subsubsection{Micro-scale topology}
The most intuitive nodal metric in classical network theory is the \textit{node degree}, which measures the actual number of links that a node share with the others.
The equivalent measure in multiplex networks is the so called \textit{overlapping} degree or strength, $o_i$, which simply sums the weighted degrees of node $i$ across all layers. 

Another popular metric to measure how the degrees of node $i$ are arranged across all layers is the \textit{multiplex participation coefficient} \cite{battiston2014}:
\begin{align}
	\label{mplex_participation}
    p_i = \frac{M}{M-1}[1-\sum_\alpha^M (\frac{k_i^\alpha}{o_i})^2],
\end{align}
where $k_i^\alpha$ is the degree of node $i$ at layer $\alpha$. When $p_i = 0$, the node's links are all concentrated in one layer; when $p_i = 1$, they are uniformly distributed across layers.

Triads of interconnected nodes, also called \textit{triangles}, are simple configurations supporting transitivity, clustering and information segregation in the network \citep{newman_networks_2010}.
Locally, this tendency is quantified via the \textit{clustering coefficient}, which measures the proportion of nodes linked to a given node that are also linked together \citep{watts1998}.
A relatively straigtforward extension is the \textit{multiplex clustering coefficient} \citep{cozzo_structure_2015}:

\begin{align}
\label{mplex_cluster}
    c_{i} = \frac{\sum_{\alpha}\sum_{\beta\neq \alpha} \sum_{j\neq i,m\neq i}a_{ij}^{\alpha}a_{jm}^{\beta}a_{mi}^{\alpha}}{(M-1)\sum_{\alpha}\sum_{j\neq i,m\neq i}a_{ij}^\alpha a_{mi}^\alpha},
\end{align}
which takes into account the possibility to form triangles by means of links belonging to two different layers.

These metrics determine which are the most \textit{central} nodes in the network. In general there are many ways of defining the centrality of a node. For example, based on the computation of shortest paths, the \textit{betweenness} of a node measures its tendency to connect topologically distant parts of the network \citep{freeman1977}. 
The extension to multiplex networks is the so-called \textit{overlapping betweenness centrality} which reads \citep{yu_selective_2017}: 

\begin{align}
\label{eq:nodal_betweeness}
 b_i = \frac{1}{(N-1)(N-2)}\sum_{\alpha} \sum_{s, s\neq t} \sum_{t, t\neq i} \frac{\sigma_{st}^\alpha(i)}{\sigma_{st}^\alpha},
\end{align}
where $\sigma_{st}^\alpha(i)$ is the number of shortest paths from node $s$ to $t$ passing through node $i$ in layer $\alpha$ and $\sigma_{st}^\alpha$ is the total number of shortest paths between node $s$ and $t$ in layer $\alpha$.

Another celebrated centrality measure is the \mbox{PageRank} centrality, initially introduced in Google web search engines \cite{brin1998}. 
PageRank centrality can be roughly thought of as the fraction of time a random walker spends visiting a node traveling through the links of the network. In multiplex networks, random walkers have the possibility to jump to adjacent nodes and teleport to nodes in other layers, according to a modified version of the transition probability (see \citet{halu2013} and \citet{de_domenico2015} for more details).

\subsubsection{Meso-scale topology}

Network \textit{motifs} are recurrent connection patterns involving few nodes and are therefore easily interpretable. They constitute the basic building blocks of a complex system architecture, coding for essential biological functions such as autoregulation, cascades and feed-forward loops \citep{milo2002,sporns2004,fabrizio2008}.

When dealing with multiplex networks, motifs can be formed by edges belonging to different layers \citep{battiston_multilayer_2017}. 
Hence, the total number of possible configurations depends on the number of layers but also on the type of interaction, e.g., negative or positive.
In these cases, $Z$-scores are typically used to determine the statistical abundance of a multiplex motif $G$ according to the following formula:

\begin{align}
Z(G) = \frac{F(G) - \bar{F}_R(G)}{S_R(G)},
\end{align}
where $F$ is the occurrence frequency of a given multiplex motif, while $\bar{F}_R(G)$ and $S_R(G)$ are respectively the mean frequency and its standard deviation obtained from a set of equivalent random multiplex graphs \textit{R}.
Alternatively, frequency coherent subgraphs can also be extracted by counting their abundance in a set of multiplex brain networks corresponding, for example, to different individuals \citep{huang_coherent_2020}.

The tendency of a network to form distinct groups, or clusters, of many nodes is an important prerequisite for the modularity of the system and its ability to process information in a segregated manner \citep{fortunato2010}.

Also known as communities, their detection is non trivial as one has to find an optimal separation that maximizes the number of links within-group and minimizes the between-group connection density \citep{newman2006}.

In the case of multiplex networks the definition of modularity incorporates the relation between different layers and partitions all the layers simultaneously \citep{mucha2010}: 

\begin{align}
\label{eq:mplex_modularity}
    Q = \frac{1}{2l}\sum_{ij\alpha \beta} [(a_{ij}^\alpha -\gamma_{\alpha}\frac{k_i^ \alpha k_j^\alpha}{2l^\alpha})\delta_{\alpha \beta} + \delta_{ij}H_{ij \beta}] \delta_{g_{i \alpha},g_{j \beta}}, 
\end{align}
where $l$ is the total number of links in the multilayer, $\gamma_{\alpha}$ sets the granularity of the community structure in each layer, $l^\alpha$ is the total number of edges in layer $\alpha$, \textit{$H_{ij\beta}$} is a parameter that tunes the consistency of communities across layers and $\delta_{{g_{i \alpha}},g_{j \beta}}=1$ when node $i$ in layer $\alpha$ and node $j$ in layer $\beta$ belong to the same community, and zero otherwise. Maximization of $Q$ is finally obtained via heuristic methods and gives an optimal network partition for each layer \citep{blondel_fast_2008}.

In temporal networks, nodal metrics reflecting mesoscale network properties can be defined by measuring, for example, the \textit{node flexibility}, i.e. the average number of times that a node changes community assignment across layers \citep{bassett_dynamic_2011}.

A peculiar network partition consists in separating the network in a \textit{core} of tightly connected nodes, and a \textit{periphery} made by the remaining weakly connected nodes \citep{borgati2000}. 
Similarly to a \textit{rich-club } \citep{colizza2006}, the presence of a core is crucial for the efficient integration of information between remote parts of the network \citep{rombach2014,verma2016,zhang_newman2015,csermely2013}.

\citet{battiston_multiplex_2018} introduced a fast core-periphery detection algorithm for multiplex networks .
Based on local information \citep{mondragon2015}, the method first defines a multiplex \textit{richness} of a node by combining its degrees in each layer. Nodes are then ranked according to their multiplex richness values and the core-periphery separation is given by the optimal rank \citep{juliana2021}:

\begin{align}
r^* = \argmax(\mu_r^+)_r,
\end{align}
where $\mu_r^+$ is the richness obtained when considering only the links of the node ranked in position \textit{r} towards nodes with higher ranks.

In the case of weighted multiplexes, the coreness of a node is given by  the number of times it belongs to the core after filtering the network with a range of different threshold values.

\subsubsection{Macro-scale topology}

Large-scale properties of complex networks are often derived by aggregating information at smaller topological scales.
For example, the \textit{global-efficiency} of a network, derived from the length of its shortest paths, quantifies the ability to integrate information from topologically distant nodes by means of a scalar number \citep{latora2001}. In a multiplex network, a straightforward extension consists in computing the shortest paths across layers.

Based on topological distances, one can also quantify the global tendency of a multiplex network to form highly clustered and efficient groups via the \textit{overlapping local-efficiency} \citep{latora2001,yu_supplementary_2017}:

\begin{align}
    E_{loc} = \frac{1}{N(N-1)}\sum_{\alpha}\sum_{i, i \neq j \in G_i }\frac{1}{k_i^\alpha(k_i^\alpha-1)}\frac{1}{d^\alpha(i,j)},
\end{align}
where $G_i$ is a sub-graph containing the neighbors of node \textit{i} and $d^\alpha(i,j)$ is the length of the shortest path between node \textit{i} and \textit{j} at layer $\alpha$. 



\citet{tang_small-world_2010} extended the concept of topological distance to temporal networks by introducing the \textit{characteristic temporal path length $L$}, which measures the formation of shortest paths across consecutive layers.

Same authors also introduced a metric to quantify the probability that the neighbor set of a node that is present at time $t$ is also present at time $t+1$.
By averaging over all the nodes they eventually defined the \textit{temporal-correlation coefficient C} as:

\begin{align}
C = \frac{1}{N(M-1)}\sum_{i=1}^N \sum_{t=1}^{M-1}\frac{\sum_j a_{ij}^t a_{ij}^{t+1}}{\sqrt{(\sum_j a_{ij}^{t})(\sum_j a_{ij}^{t+1})}}
\end{align}

Together, the last two global metrics measure how the system information is respectively integrated and segregated over time and can be used to assess the small-world properties of time-varying networks \citep{tang_small-world_2010}.

In graph theory, the \textit{Laplacian} matrix has many useful implications in real networks, from denoising to low-dimensional embedding \citep{merris1994}. The second smallest eigenvalue of the Laplacian, also called \textit{algebraic connectivity }~($\lambda_2$) plays an important role since it informs on several important properties of a network such as community structure, synchronization, diffusion and resilience \citep{fortunato2010}.

In a full multilayer network, $\lambda_2$ is calculated from the associated supra-Laplacian matrix, whose elements are defined as:

\begin{equation}
\mathcal{L}_{ij}^{\alpha \beta} = 
\begin{cases}
\mu_i^{\alpha} -a_{ii}^{\alpha}  & \text{\small{, if \textit{i = j, $\alpha=\beta$}}} \\
\quad \quad -a_{ij}^{\alpha \beta}  & \text{\small{, otherwise}}\\
\end{cases}
\end{equation}
,where $\mu_i^\alpha$ is the is the total number of links incoming to node \textit{i} at layer $\alpha$.

In multilayer networks, $\lambda_2$ is sensitive to the amount of intra- and inter-layer connectivity, and typically quantifies the integration/segregation balance among layers from a dynamical perspective \citep{gomez2013, radicchi_abrupt_2013}.
Notably, $\lambda_2$ exhibits a phase transition when increasing the interlayer connection intensity, from layers being independent/segregated to a high overall dependence/integration \cite{radicchi_abrupt_2013}. 

\section{Multilayer brain networks}
\subsection{Common types of multilayer brain networks}
\label{sec:type_multilayer}

Up to date multilayer brain networks have been mostly derived from experimental neuroimaging data in humans, with nodes representing the same entities, i.e. brain areas across layers.
Multiplex networks represent the easiest way to bridge brain connectivity at different levels, as one does not have to explicitly infer interlayer connections. In this situation, interlayer links only virtually connect the replica nodes and the associated meaning is basically the one of identity between the same nodes across layers (\textbf{Fig.~\ref{fig:type_multilayer}b})\citep{battiston2014}.

This type of representation has been largely used to describe multimodal brain networks, whose different layers may contain structural and functional connectivity \citep{battiston_multiplex_2018,lim_discordant_2019,simas_algebraic_2015}, as well as interactions at different signal frequencies \citep{guillon_loss_2017,yu_selective_2017,domenico_mapping_2016}. 
A common situation when dealing with multimodal networks is that the nodes might not correspond to the same entity in their native space. This is for example the case of brain networks derived from fMRI and EEG signals, where nodes correspond respectively to image voxels and scalp sensors. 
To overcome this issue, advanced image and signal processing tools are used beforehand for projecting the native signals into the nodes of a common anatomical brain space, typically extracted from the structural MRIs of a subject's head \citep{baillet2001,michel2004,grech_review_2008}.
Multiplex networks have been also adopted to describe temporal brain networks, i.e. networks whose topology is changing over time \citep{bassett_dynamic_2011,pedersen_multilayer_2018,braun_dynamic_2015}. In this case, each layer corresponds to a specific point, or instance, in time and only the replica nodes of temporally adjacent layers are interconnected according to a \textit{Markovian} rule (\textbf{Fig.~\ref{fig:type_multilayer}c}). Unlike multimodal brain networks, the layers of a time-varying brain network do not correspond to different spatial or temporal/frequency scales, but they typically capture the dynamic network evolution within a fixed time resolution. This is typically in the order of milliseconds for motor behavior, minutes/hours for human learning, or years for aging as well as for neurodegenerative diseases. 

Full multilayer network representations, containing both intra-layer and inter-layer nontrivial connectivity, have been mostly adopted to characterize brain signal interactions within and between different oscillation frequencies (\textbf{Fig.~\ref{fig:type_multilayer}a}) \citep{tewarie_integrating_2016,buldu_frequency-based_2018,tewarie_interlayer_2021}. 
This representation is particularly useful for functional brain networks with a broad frequency content, such as in those obtained from electrophysiology, EEG or MEG signals.
Although less frequent than multiplexes, this type of representation has a great potential for characterizing whole brain cross-frequency coupling, which has been recently shown to be crucial for many cognitive and pathological mental states \citep{jirsa2013}.

We finally stress out that regardless of the type of construction, the resulting multilayer networks---either multiplex or full---generally exhibit higher-order properties that cannot be captured or resumed by simply aggregating information from different layers \citep{boccaletti2014,kivela2014}. 

\subsection{Multilayer brain networks are more than the sum of their layers}
\label{sub:multilayer_method}
\begin{figure}[t]
\includegraphics[scale=0.4]{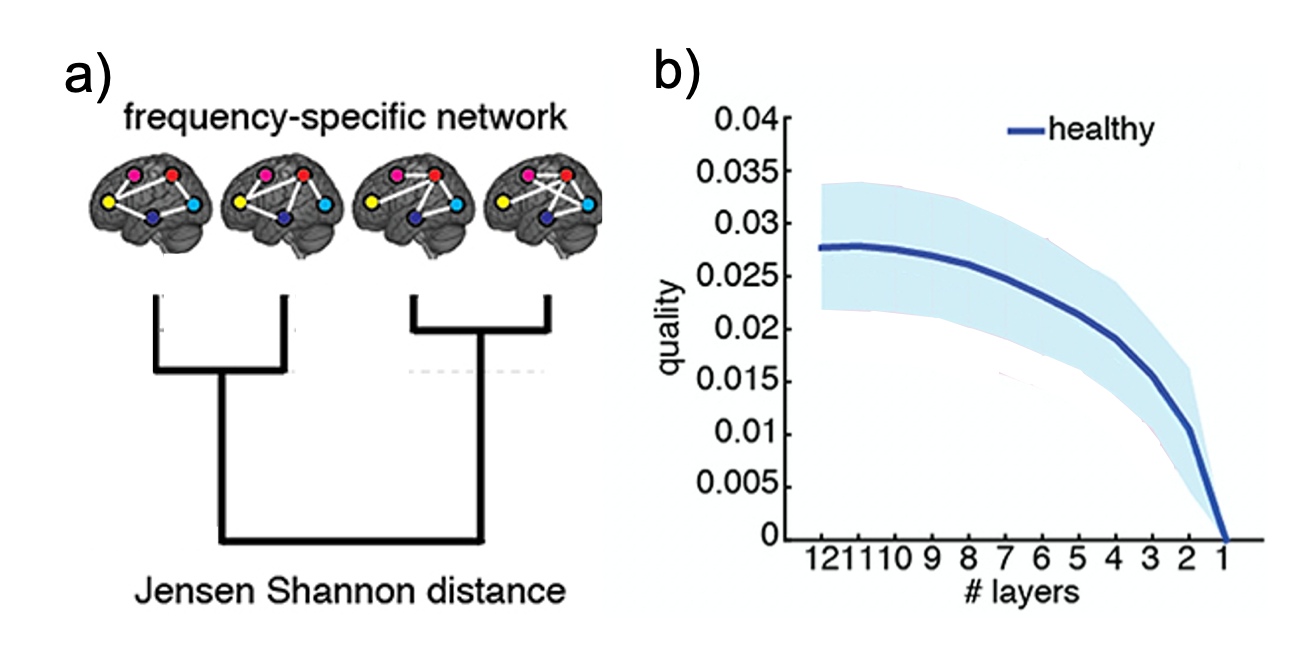}
\caption{\label{fig:reduciblity} \textbf{Structural reducibility of multifrequency brain networks.} 
Panel a) For each combination of layers a quality function measures the amount of new information added with respect to an equivalent single-layer model.
Panel b) Median values of quality function obtained from fMRI multifrequency brain networks in healthy subjects. Shaded areas indicate the standard deviation around each value.
Pictures and captions adapted from \citet{domenico_mapping_2016}.}
\end{figure}

\begin{figure}[b]
\includegraphics[scale=0.35]{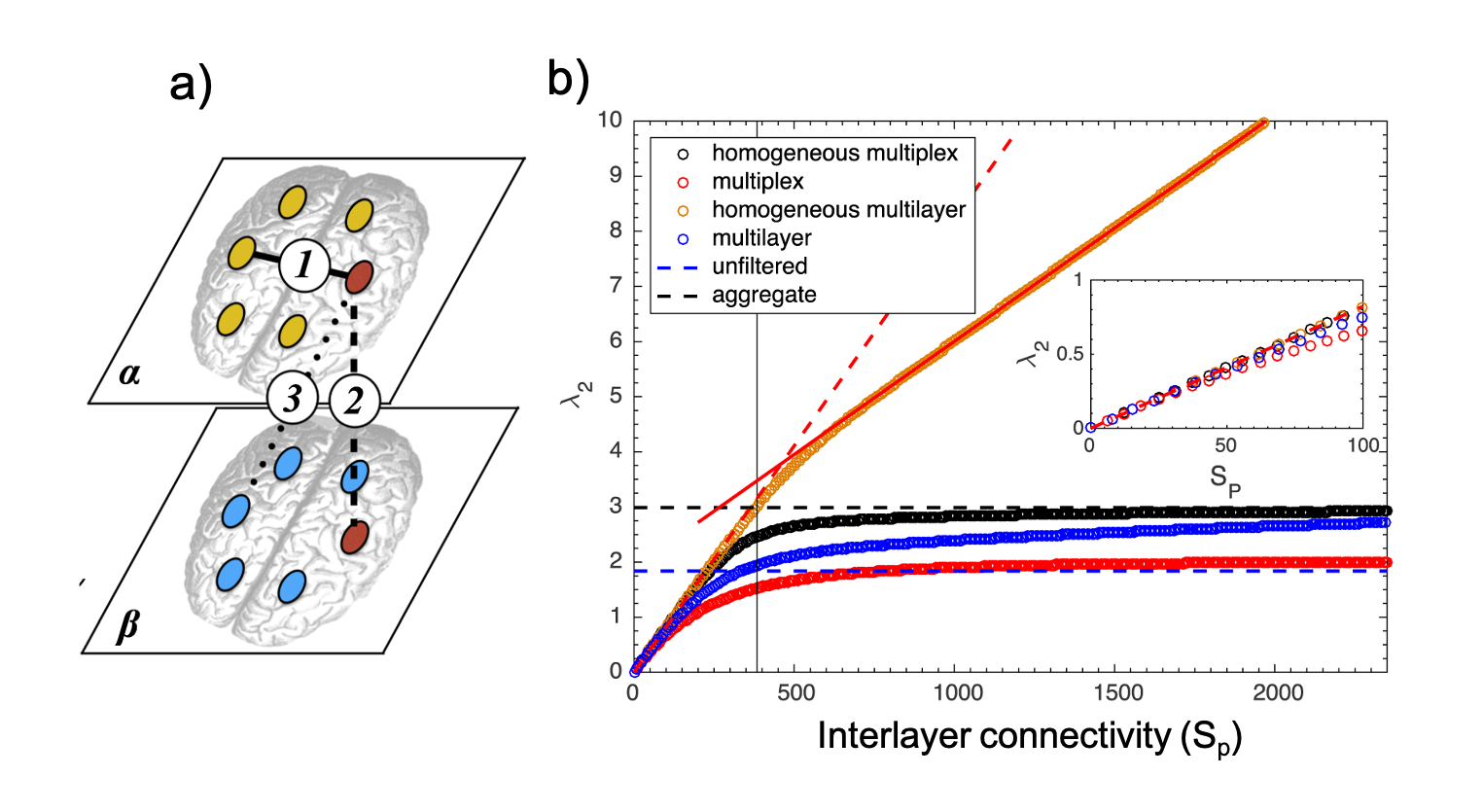}
\caption{\label{fig:buldu_porter}\textbf{Emergent properties in full multilayer brain networks} Panel a) Intralayer and interlayer edges in the multifrequency MEG network. 1-edge between regions at the same frequency; 2-edge of the same area between different frequency bands; 3-edge between different nodes at different frequency bands.
Panel b) Algebraic connectivity $\lambda_2$ as a function of the total interlayer connectivity ($S_p$).  The vertical solid line corresponds to the actual value of interlayer connectivity, i.e., without modifying their weights. Pictures and captions adapted from \citet{buldu_frequency-based_2018}.}
\end{figure}

Multilayer networks give richer description than standard network approaches, but do they really represent a step forward into the modeling of brain organization? Why aggregating layers is not enough? Are all layers necessary to capture the main organizational properties?
\citet{de_domenico_structural_2015} addressed these questions by introducing a \textit{structural reducibility} approach to maximize the quantity of non-redundant topological information between the layers of a multiplex network with respect to its aggregated counterpart (\textbf{Fig.~\ref{fig:reduciblity}a}). For a large spectrum of networks, from protein-protein interactions to social networks, structural reducibility showed that the best configuration in terms of distinguishability is not necessarily the one with the highest number of layers \citep{de_domenico_structural_2015}. 
On the contrary, \citet{domenico_mapping_2016} showed that multifrequency brain networks derived from fMRI signals were not easily reducible since all the layers brought some non-redundant topological information (\textbf{Fig.~\ref{fig:reduciblity}b}).
This result implies that even if fMRI oscillations are under-represented in higher frequencies, their broad interaction remains crucial for a correct brain functioning.
We show in the next sections that this result extends quite generally and can be used to better diagnose brain diseases (see Sec.~\ref{sec:brain_org} and Sec.~\ref{sec:brain_disorder}).

While most research has focused on multiplex brain networks, a better understanding of the emerging properties in full multilayer brain networks still remains to be elucidated. 
\citet{buldu_frequency-based_2018} addressed these aspects by studying the difference between frequency-based multiplexes and full multilayers derived from MEG brain signals (\textbf{Fig.~\ref{fig:buldu_porter}a}).
By evaluating the algebraic connectivity $\lambda_2$ (see Sec.~\ref{sub:tool}), they showed that full multilayer brain networks are close to an optimal transition point between integration and segregation of the layers. The layers in the equivalent multiplex configurations were instead more segregated and then far from this transition point \citep{radicchi_abrupt_2013}.
These results were also confirmed by extensive numerical simulations and explained by the intrinsic lower interlayer connection density of the multiplexes (\textbf{Fig.~\ref{fig:buldu_porter}b}).
Interestingly, the full multilayer $\lambda_2$ values were associated with the phase-amplitude coupling of \textit{gamma} ($30-40$ Hz) and \textit{theta} ($4-7$ Hz) brain frequency bands, confirming the crucial role of cross-frequency coupling in the study of complex brain functions and dysfunctions \citep{canolty2006,aru2015}.
These findings point out the importance of considering previously unappreciated cross-layer interactions to explain the emergent properties of brain organization.


\subsection{Filtering spurious links in multilayer brain networks}

It's important to remind that brain connectivity networks are estimated from experimental data.
This necessarily implies the presence of spurious connections, often among the weakest ones, due to the statistical uncertainty associated with the connectivity estimator and/or to the presence of signal artifacts during the experiment \citep{fabrizio2014,korhonen2021}.
For example, head motions are known to abnormally increase short-range connectivity, thus altering the original topology of the network as well as its connection intensity, i.e. the sum of the actual links' weights \citep{lydon-staley_evaluation_2019}. This is particularly relevant as the topological properties of a network strongly depend on the number and weights of the existing edges \citep{fabrizio2017,mandke_comparing_2018}.
As a result of the construction process, multilayer brain networks are also influenced by such noise, which might alter the true association between the multiscale brain network organizational properties and the subject's characteristics and behavior.

To mitigate the presence of unwanted alterations in the estimated links, two main strategies have been so far adopted following what has been done in standard network analysis.
The first approach consists in manipulating the brain signals, while the second one operates directly on the connectivity matrices.
\citet{lydon-staley_evaluation_2019} used the first approach to silence the effects of head motion on the recorded brain signals and in turn on the estimated brain network.
They tested different signal denoising strategies, mainly based on regression and source separation techniques \citep{regression2002} on temporal brain multiplexes constructed from fMRI data. 
Specifically, they evaluated their ability in attenuating the nuisance effects on several network metrics, such as multiplex modularity and node flexibility (cf.~Sec.~\ref{sub:tool}).
Despite some variability, the obtained results suggested that regression-based approaches outperform source separation-based techniques, possibly due to their ability to explicitly incorporate the nuisance variables in the denoising process \citep{lydon-staley_evaluation_2019}.

The second approach consists in filtering the network's links. This is typically achieved by fixing a threshold either on the percentage of strongest edges to retain or on their weights. Depending on the threshold value the resulting networks might have different densities and/or intensity.

\citet{mandke_comparing_2018} evaluated the impact of network filtering on several topological properties such as multiplex PageRank (cf.~Sec.~\ref{sub:tool}), multiplex modularity (Eq.~\ref{eq:mplex_modularity}) and participation coefficient (Eq.~\ref{mplex_participation}). 
Specifically, they tested several filtering criteria, e.g., spanning tree (MST) \citep{kruskal}, efficiency cost optimization (ECO) \citep{fabrizio2017}, singular value decomposition (SVD) \cite{svd} applied to each single layer separately, or adapted to the whole multiplex.

By using both synthetic and neuroimaging-derived multiplex networks, results indicated that SVD techniques lead to multilayer network properties that are quite robust to changes in connection density/intensity.
MST and ECO techniques were instead effective only when filtering each layer separately, and therefore useful when dealing with multimodal brain networks, where layers are estimated from different type of data and the nature of the interlayer links cannot be straightforwardly established.

Note however, that these results have been obtained for multiplexes and the extension to full multilayer networks still remains to be investigated.

\section{Multilayer network properties of brain organization}
\label{sec:brain_org}
\subsection{Structure-function relationship}

\begin{figure*}
\includegraphics[scale=0.45]{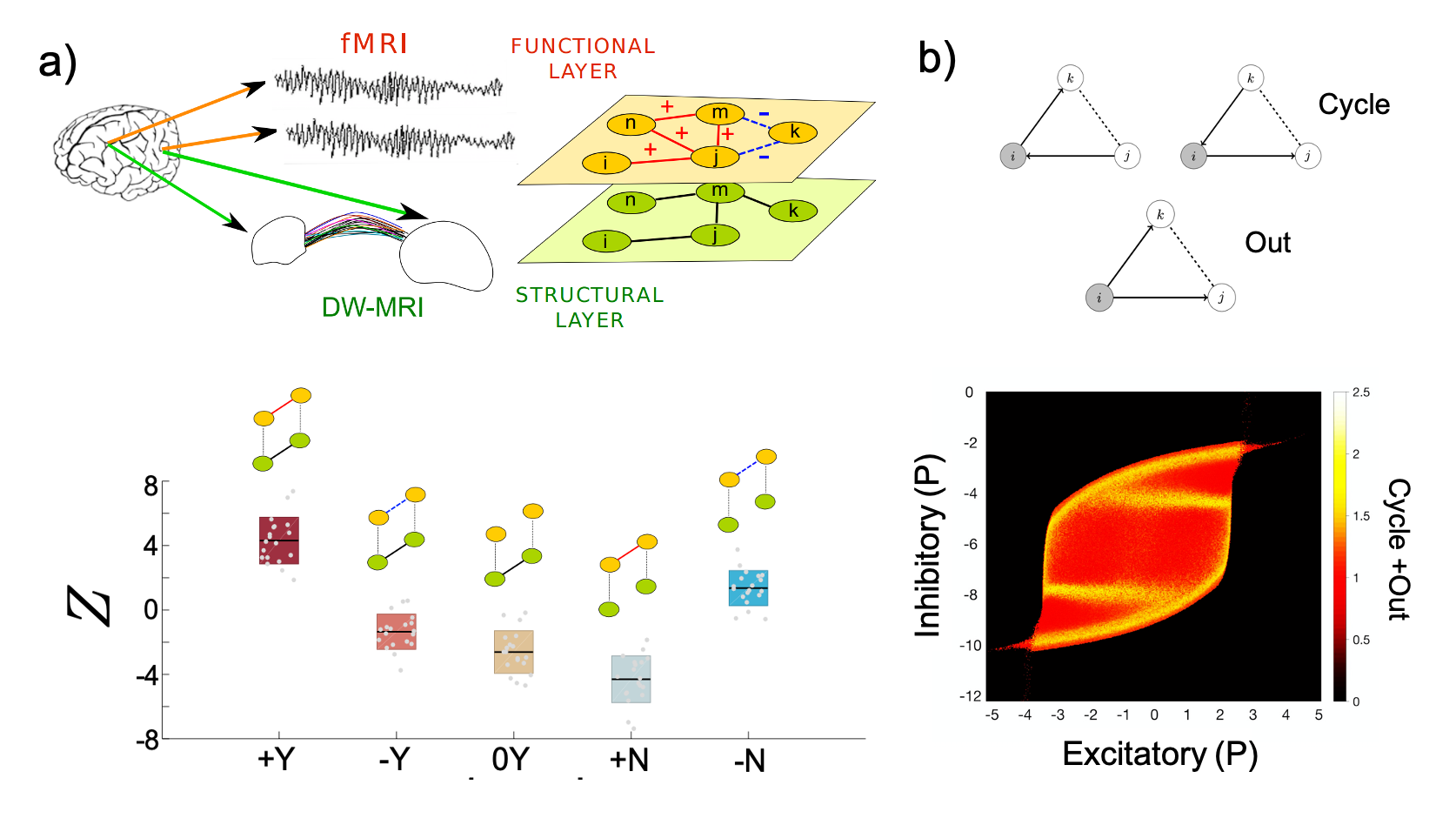}
\caption{\label{fig:mplex_motifs}  \textbf{Multiplex motif analysis of multimodal brain networks.} 
Panel a) Structural-functional 2-layer brain network. Interlayer links between replica nodes are omitted for the sake of visibility. Five nontrivial multiplex motifs of two nodes are possible based on the type of connectivity in the DTI structural layer and in the fMRI functional layer. The Z-scores show the motifs that are overrepresented and underrepresented as compared to equivalent random networks. 
Panel b) Patterns of multiplex triangles comprising directed structural tuples (solid connections) closed by a functional edge (dashed connections). The overall motif counts normalized by equivalent random multiplexes are illustrated as a function of basal activation parameters P and Q of the Wilson-Cowan model. Pictures and captions is adapted from \citet{battiston_multilayer_2017}, with permission of AIP Publishing (panel a), and  \citet{crofts_structure-function_2016} (panel b).}
\end{figure*}

Both structural and functional brain organization are crucial determinants of complex neural phenomena such as cognition, perception, and consciousness \citep{friston2013}.
An important question in modern neuroscience is how structural and functional connectivity are related to each other, and how such putative interaction can better our understanding of the brain organization.
Recent studies using both model-based and data-driven approaches have for example demonstrated that connectivity at functional level could be in part predicted by the structural one, and that this prediction could explain several complex dynamics of brain functioning, from resting states to task-based and pathological conditions \citep{arico_brain_2021,suarez2020,HANSEN2015,friston2013}.

But what are the higher-order topological properties of the multilayer network composed of both structural and functional layers and how these contribute to describe brain anatomo-functional organization?
To address these questions, \citet{battiston_multilayer_2017} first investigated the presence of simple connection motifs (cf.~Sec.~\ref{sub:tool}) forming across the layers of a DTI-fMRI multiplex network. 
They found that motifs comprising both structural and positively correlated functional links are overabundant in the human brain (\textbf{Fig.~\ref{fig:mplex_motifs}a}). This confirms that the presence of an anatomical connection is likely to induce a synchronized activity between the corresponding brain regions \citep{SKUDLARSKI2008}. However, other significant configurations were reported including the presence of triangles in the functional layer with no support in the structural one.
Overall these results indicated that intrinsic functional organization of the brain is non-trivially constrained by the underlying anatomical network \citep{SKUDLARSKI2008} and cannot be solely explained by it.

Down the line, \citet{ashourvan_multi-scale_2019} investigated the multilayer modularity of DTI-fMRI multiplex networks.
Main results showed that the structural layer is mostly dominating the community structure of the multiplex over a broad range of topological scales explored by varying the granularity parameter $\gamma$  (Eq.~\ref{eq:mplex_modularity}).
Notably, the communities of the structural layer tended to spatially overlap with the cytoarchitectonic brain organization and were highly consistent across individuals. Instead, the communities of the functional layer were more heterogeneously distributed and less consistent across subjects, reflecting the dynamic repertoire of the brain functions \citep{ghosh2008,hadriche2013}. 


By looking at DTI-fMRI multiplex networks, \citet{lim_discordant_2019} measured to what extent nodes with similar overlapping degrees tended to "wire" together, a property often referred as to assortativity.
Results indicated that multimodal brain networks have a propensity to be assortative, which translates into an overall ability to facilitate system dynamics and resilience to random attacks (e.g., node removal) \cite{boccaletti2014}. This evidence resolved the assortative/disassortative dichotomy  previously observed with single-layer analysis of structural/functional brain networks.
Notably, such multilayer assortativity resulted from a nontrivial structure/function interplay and pointed out a novel organizational mechanism optimally balancing the resilience to damages and restrainability of their effects.

Modeling the emergence of large-scale brain dynamics from microscale neuronal interactions is crucial for a mechanistic understanding of neural multiscale organization. 
An early study by \citet{zhou2007} proposed a computational model based on the structural connectome of the cat cortex. 
By parametrizing the coupling between several Fitzhugh-Nagumo oscillators according to the available connectome, they simulated the ongoing activity in each region, and estimated the interareal functional connections via Pearson's correlation \citep{fitzhugh}. 
By means of this simple model, the Authors showed that a weak coupling parameter was sufficient to generate biologically plausible macroscale activity, with functional connectivity patterns mostly overlapping the modular organization of the structural network. 

\citet{crofts_structure-function_2016} used a similar approach based on the structural connectome of a macaque cortex and Wilson-Cowan neuronal models \citep{WILSON1972}.
More relevant to this Colloquium, they analyzed the behavior of multiplex clustering patterns (such as in Eq.~\ref{mplex_cluster}) in the structural-functional networks as a function of two model parameters, i.e. one tuning the input to excitatory neurons, and the other one modulating the input to the inhibitory ones.
Specifically, they defined multiplex clustering indices to quantify the presence of functional links associated with common drivers in the structural layer.
Main results showed that such quantities were maximal at the boundaries of the phase transition, from steady-state to oscillatory dynamics, as well as in other regions of the parameter space (\textbf{Fig.~\ref{fig:mplex_motifs}b}).
Differently from previous results on single-layer analysis, this nontrivial behavior suggested that the system criticality does not only depend on the structure-functional interplay of the brain network, but also on the type of ongoing dynamics.

At the level of single neuron, \citet{bentley_multilayer_2016} proposed a multiplex approach to represent synaptic connections (structural) as well as extrasynaptic signaling interactions (functional) inferred from gene expression data of the \textit{C.~Elegans} worm.
Despite the low degree of overlap between the synaptic and extrasynaptic connectomes, Authors found highly significant multiplex motifs (similar to the ones in Sec.~\ref{sub:tool}), pinpointing locations in the network where aminergic and neuropeptide signalling modulate synaptic activity.
The presence of directed monoamine interactions and reciprocal synaptic connections was particularly significant among specific neurons implicated in learning, memory and motor functions.
These results support the evidence that the structural/functional interplay is crucial to better understand the communication pathways between different parts of the \textit{C.~Elegans} nervous system.

In this direction, \citet{maertens_multilayer_2021} identified the shortest paths from touch sensory neurons to motor neurons allowing information flowing across different type of neurotransmitters and neuropeptides layers.
By applying a time-delayed feedback control on the identified neurons, the Authors could eventually reproduce the typical \textit{C.~Elegans} locomotion, and characterize the neuromuscular multilayer connectivity mechanisms associated with the central pattern generator \citep{fouad2019,cpg2014}.

Multilayer network theory has just started to provide new tools and insights into the complex interplay of the brain structure and function. Several issues remain to be explored such as how to establish interlayer connections \citep{tewarie_interlayer_2021} or incorporate multilayer network mechanisms in the laws modeling the large-scale neuronal dynamics \citep{HANSEN2015}.

\subsection{Information segregation and integration}
\label{subsec:segregation}

\begin{figure}

\includegraphics[scale=0.44]{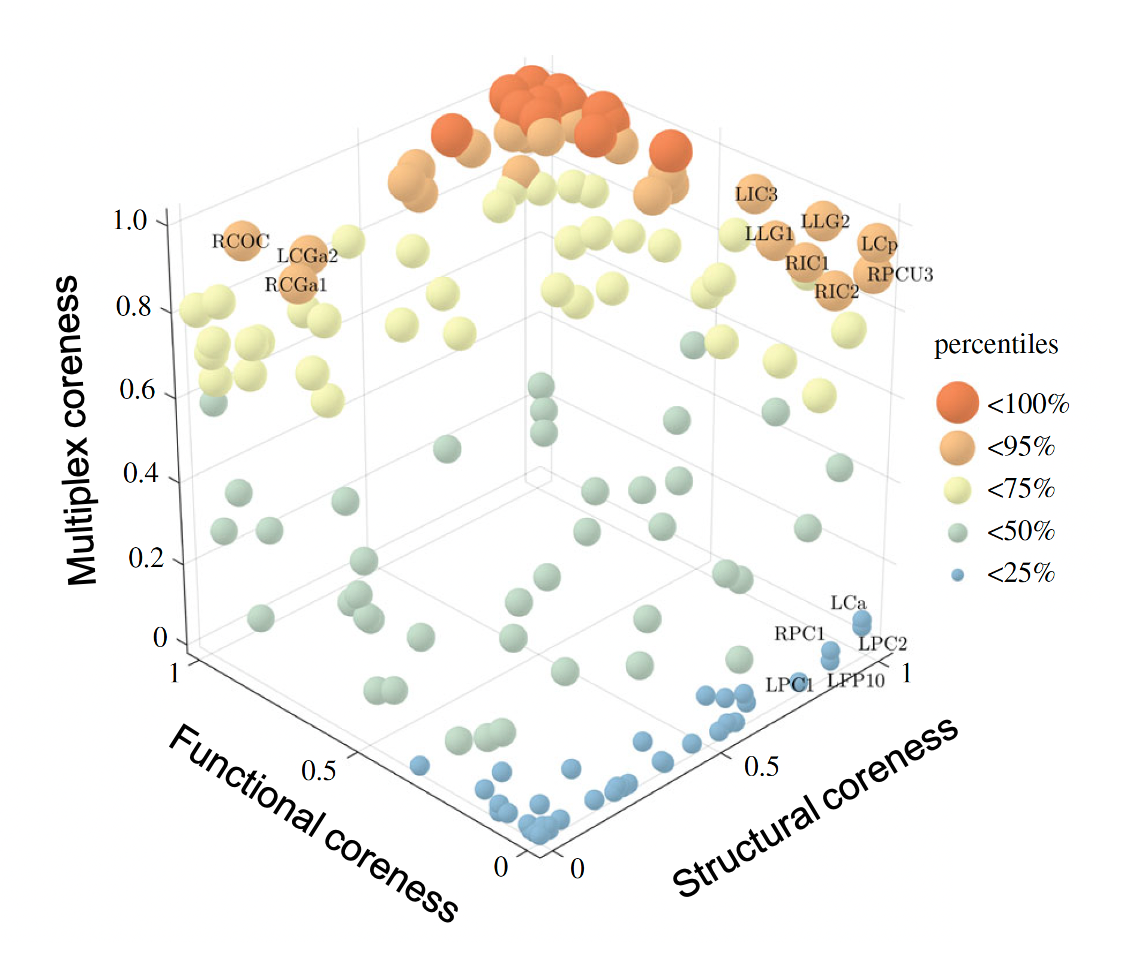}
\caption{\label{fig:core_periphery} \textbf{Multiplex core-periphery structure of the human connectome.}
Scatter plot of multiplex coreness against single-layer corenesses obtained from structural (DTI) and functional (fMRI) layers. Labels indicate brain areas whose multiplex coreness cannot be predicted by looking at the coreness values in the respective structural and functional layer.
Picture and caption adapted from \citet{battiston_multiplex_2018}, republished with permission of The Royal Society (U.K.); permission conveyed through Copyright Clearance Center Inc.
}
\end{figure}

\begin{figure}
\includegraphics[scale=0.51]{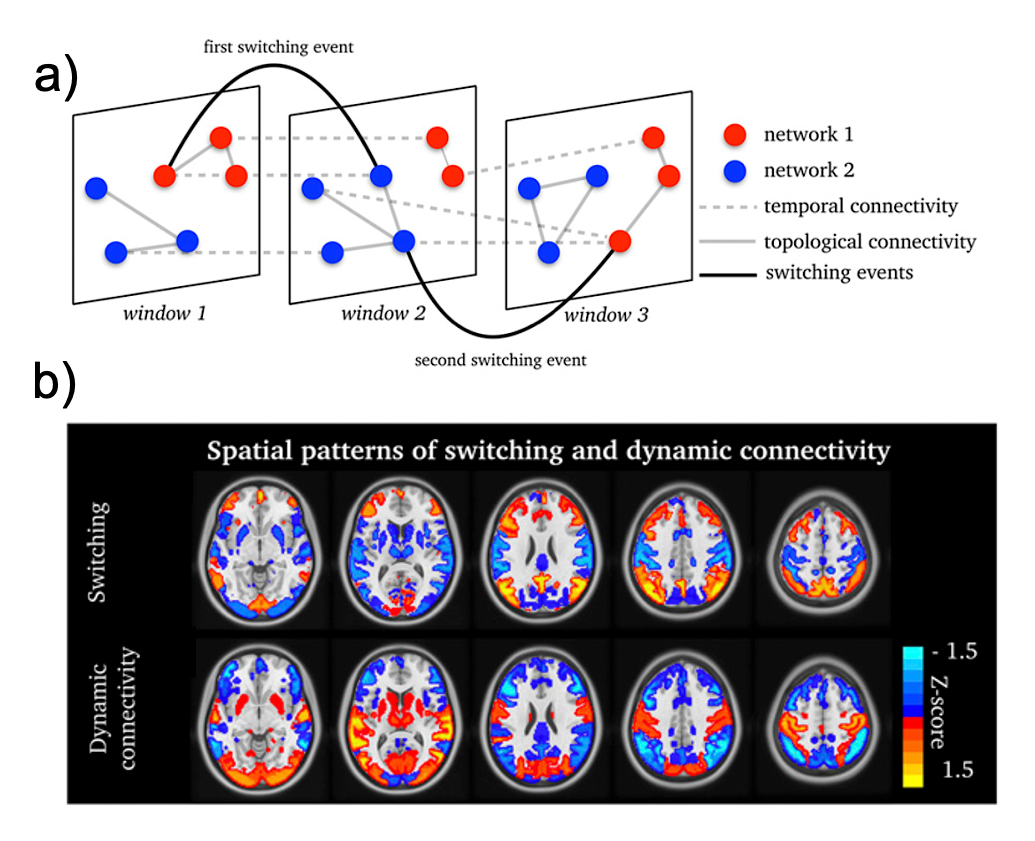}
\caption{\label{fig:switching} \textbf{Temporal network flexibility correlates with brain perfomance.}
Panel a) An overview of network switching (or flexibility) in a temporal network. Red and blue colors identify the nodes belonging to two different communities according to the multilayer network modularity metric. 
Panel b) Brain maps of switching rate and dynamic fMRI connectivity. Values were normalized into z-scores so ensure both connectivity dynamics and switching values were scaled equally. Pictures and captions adapted from \citet{pedersen_multilayer_2018}}.

\end{figure}

Clustering and shortest paths are general concepts in complex systems that are both essential for efficient organization of many real-world networks \citep{watts1998,latora2001}. These concepts reconcile two long-standing opposed views of the brain functioning. On one hand, phrenology-based theories, which associated different cognitive tasks with segregated brain regions \citep{Kanwisher11163}. On the other hand, global workspace theories, which instead hypothesize the necessity of interareal integration of information to realize the very same tasks \cite{DEHAENE2001}.
Notably, network science has provided the tools to quantify network segregation and integration by demonstrating respectively the presence of many clustered connections and few shortest paths between areas.
More recently, integration in the brain has been revisited and hypothesized to be determined by the presence of few core hubs in the network, and not directly by shortest paths \citep{obando2017,deco_rethinking_2015}.
By considering  multilayer brain networks, segregation and integration becomes a joint property of both nodes and layers, thus providing information about higher-order phenomena such as cross-frequency coupling \citep{jirsa2013}, multimodal information \citep{garces2015} and temporal evolution \citep{HUTCHISON2013}.


\citet{tewarie_integrating_2016} investigated information segregation and integration in MEG full multifrequency brain networks.
They first observed the presence of strong dependencies between intra- and interlayer connectivity. 
By decomposing the multilayers into representative connectivity structures, or ``eigenmodes", they demonstrated that the overall amount of interlayer connectivity was associated with the second eigenmode, containing specific fronto-occipital network components common to all frequencies.
In addition, they compared the empirical MEG multifrequency networks with those obtained from large-scale signals simulated with a thalamo-cortical model \citep{robinson2001,robinson2002}. 
By increasing the model structural coupling parameter, the Authors reported a progressive increase in the resulting functional interlayer connectivity.
Notably, real MEG multilayer networks maximally fit the model at the transition point of such increment, suggesting an optimal balance between segregation and integration of information between different frequency bands.

As for multimodal connectivity, \citet{battiston_multiplex_2018} investigated the associated integration properties by evaluating the core-periphery structure of DTI-fMRI multiplex networks. 
They specifically calculated the multiplex coreness (cf.~Sec.~\ref{sub:tool}), which integrates information from different layers and provide a possibly more accurate characterization of the mesoscale brain network properties.
Compared to single-layer analysis, results identified new core areas in the sensorimotor region of the brain that are key components of the so-called default mode network (DMN), i.e. a set of brain regions that is active when a person is not focused on the outside world \cite{raichle2001}.
Besides, results excluded previously established areas in the frontal region, whose belonging to the core system was still debated \citep{hagman2008}.
By including structural (DTI) and functional (fMRI) network information, these findings offered a new enriched description of the integration properties of the human connectome's core (\textbf{Fig.~\ref{fig:core_periphery}}).

Temporal brain networks have been previously shown to exhibit alternating periods of segregation and integration across multiple time scales, associated with the presence of "dynamical" hubs \citep{de_pasquale_dynamic_2016}, as well as state-dependent community structures \citep{al-sharoa_tensor_2019}.
To better understand the role of such transitions, \citet{pedersen_multilayer_2018} studied the multilayer network flexibility (cf.~Sec.~\ref{sub:tool}) derived from a big dataset of resting-state fMRI signals (\textbf{Fig.~\ref{fig:switching}a}).
Results showed that the node flexibility, i.e. the frequency of community switching between consecutive time layers, was particularly high in specific associative brain regions (i.e., temporal and parietal) and correlated with the entropy of the connectivity variability.
Because switching is known to increase in systems with high entropy or information load \citep{entropy_switch}, Authors eventually established the role of functional hubs for the associative cortex integrating information across differently specialized brain systems \citep{heuvel2011}.
Interestingly, these high local flexibility values occurred mainly when the brain exhibited a globally low and steady network intensity, so as to minimize the overall energetic cost associated with the integrative temporal switching (\textbf{Fig.~\ref{fig:switching}b}).

On longer time scales, \citet{malagurski_longitudinal_2020} investigated how brain segregation changes with age by using longitudinal fMRI data acquired over a $4$ years time-span.
By computing the multiplex modularity (Eq.~\ref{eq:mplex_modularity}), they showed that the global flexibility, i.e. the average node flexibility, is significantly higher in healthy elderly as compared to a temporal null model, where the brain network layers are randomly shuffled \citep{chai_functional_2016,sizemore2017}. 
Results also demonstrated that people with more segregated temporal networks tended to be more resistant to transient changes in modular allegiance \citep{harlalka2019,meunier2010,ramos2017}.
Notably, older age was related to higher temporal variability in modular organization. However, no correlations were found with cognitive behavior, such as processing speed and memory encoding.
Since flexibility is in general a good predictor of cognitive performance (cf.~Sec.~\ref{sub:learning}), further studies should include more cognitive domains, or lagged changes, to elucidate the role of age in the relation between the cognitive performance and temporal modular flexibility.

Taken together, these findings provided some concrete examples on how concepts such as segregation/integration of information can be broaden to capture multilayer brain mechanisms and provide complementary information about the system's behavior.
While most of the studies have focused on undirected connectivity, future research will be crucial to include directed links and better inform on communication pathways in neuronal systems \citep{avena-koenigsberger_communication_2018}.

\subsection{Brain organizational properties of human behavior}
\label{sub:learning}
\begin{figure}
\includegraphics[scale=0.5]{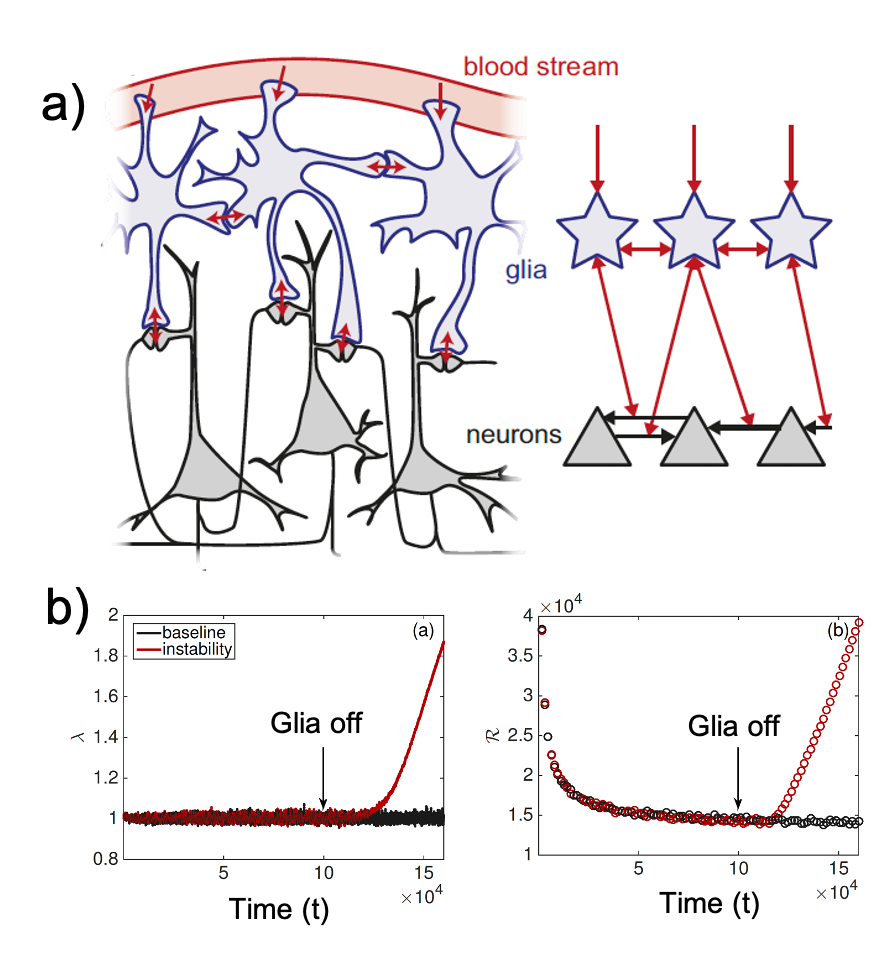}
\caption{\label{fig:glia}\textbf{Stabilization of critical dynamics in multilayer glia-neuronal networks.} Panel a) Left side: Glia cells redistribute metabolic resources from the bloodstream to neural synapses. Right side: Associated two-layer network model. Black arrows indicate neural synaptic interactions. Arrow thickness indicates synaptic strength which evolves according to spike time-dependent plasticity (STDP). Red arrows which terminate on black arrows represent the resource supply to the corresponding synapse.
Panel b) Stability analysis of the two-layer STDP model. The largest eigenvalue $\lambda$ of the neuronal network layer and the total resource $\mathcal{R}$ of all glia and synapses are illustrated as function of time. The data plotted in black correspond to a `baseline' condition. For the data plotted in red (labelled 'instability'), the initial evolution is the same as for the baseline data up until the diffusion of resources between the glial cells is turned off (vertical arrow). Pictures and captions adapted with permission from \citet{virkar_feedback_2016}. Copyright 2016 by the American Physical Society. }
\end{figure}

The results presented in the previous paragraphs aimed to quantify the intrinsic structural and functional brain organization, with no reference to any specific mental state or behavior.
Nonetheless, the brain is an extremely flexible and adaptive system, capable of altering its organization depending on endogenous and exogenous stimuli coming from the external environment (a property often referred to as \textit{plasticity}). 
In this paragraph, we present some of the most recent results showing how multilayer brain network properties change according to specific behaviors, and how those higher-order topological changes are associated with inter-subject variability.

Human learning is perhaps one of the most intriguing (yet not completely understood) neural processes with numerous implications in our daily-life \citep{zatorre_plasticity_2012, BARAK2014}. A basic question in neuroscience is how learning is acquired through \textit{Hebbian} plasticity, without leading to runaway excitation of the neural synaptic activity \citep{watt2010,miller1994,abbott2000}.
In their study, \citet{virkar_feedback_2016} proposed a mechanism for preserving stability of learning neural systems, via a 2-layer network model. The first layer contained a model neural network interconnected by synapses which undergo spike-timing dependent plasticity (STDP) \citep{feldman2012}. The second layer contained a network model of glia cells interconnected via gap junctions, which diffusively transport metabolic resources to synapses (interlayer edges) (\textbf{Fig.~\ref{fig:glia}a}).
Main results showed that, with appropriate model parameter values, the diffusive interaction between the two layers prevents runaway growth of synaptic strength, both during ongoing activity and during learning.
These findings suggested a previously unappreciated role for fast dynamic glial transport of metabolites in the feedback control stabilization of slow neural network dynamics during learning (\textbf{Fig.~\ref{fig:glia}b}). 
Notice that this is so far one of the few examples where multilayer network theory is used to model microscale neural organization across multiple temporal scales.

\begin{figure}
\includegraphics[scale=0.5]{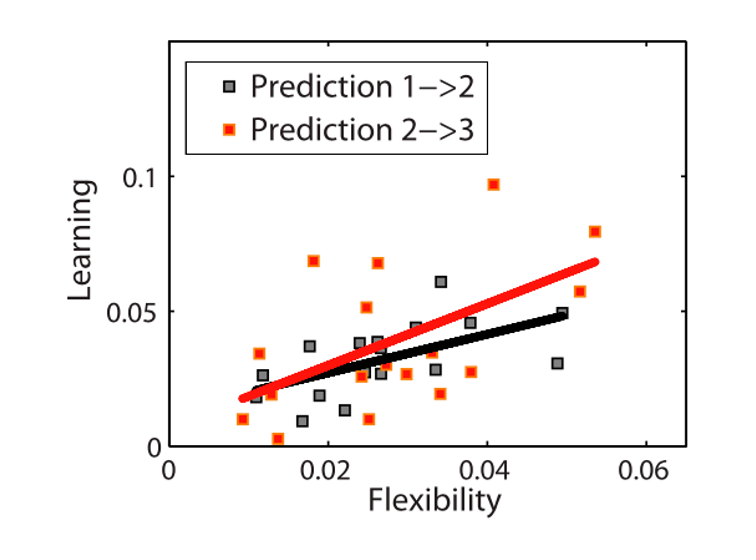}
\caption{\label{fig:learning_rate} \textbf{Temporal network flexibility predicts future learning rate. 
} Significant predictive Spearman correlations between flexibility in session 1 and learning in session 2 (black curve, $p \approx 0.001$) and between flexibility in session 2 and learning in session 3 ($p \approx 0.009$). Each point corresponds to a subject. Note that relationships between learning and fMRI network flexibility in the same experimental sessions (1 and 2) were not significant; $p > 0.13$ was obtained using permutation tests. Picture and caption adapted from \citet{bassett_dynamic_2011}.}
\end{figure}

At larger spatial scales, \citet{bassett_dynamic_2011} used a multilayer network approach to characterize human learning during a simple motor task. 
In particular, they built temporal brain networks from fMRI signals across consecutive experimental sessions.
They used the multilplex modularity (Eq.~\ref{eq:mplex_modularity}) to find long-lasting modules and found that community organization changed smoothly with time, displaying coherent temporal dependence, as in complex long-memory dynamical systems \citep{achard2008}.
Results also showed that the network flexibility changed during learning---first increasing and then decreasing---demonstrating a meaningful biological process. 
In particular, the nodal flexibility (cf.~Sec.~\ref{sub:tool}) was stronger in frontal, posterior parietal and occipital regions. Also, it predicted the relative amount of learning from one session to the following one (\textbf{Fig.~\ref{fig:learning_rate}}). These predictions could not be obtained via conventional task-related fMRI activation or standard network analysis, and confirmed the relation between network flexibility and cognitive performance.
Indeed, network flexibility has been found to correlate with several mental states, such as working memory and planning \citep{pedersen_multilayer_2018,braun_dynamic_2015}, but also with mental fatigue \citep{betzel_positive_2017} and sleep deprivation \citep{pedersen_multilayer_2018}.
At this stage, it would be interesting to elucidate whether network flexibility is an aspecific predictor of cognitive performance or it can also distinguish between different dynamic brain states.

\citet{makarov_betweenness_2018} further study the cognitive load during attentional tasks in a EEG frequency-based multiplex framework. Based on betweenness centrality (cf.~Sec.~\ref{sub:tool}), they observed an outflow of shortest paths from low frequencies toward high frequencies in the fronto-parietal regions. 
These findings suggest that cross-frequency integration of information is not only an intrinsic characteristic of the brain functioning \citep{tewarie_integrating_2016}, but it is also modulated by attentional tasks as well as by drowsiness \citep{harvy_between-frequency_2019}.

In a recent study, \citet{williamson_multilayer_2021} investigated how the brain supports expressive language function by looking at MEG multifrequency brain networks.
In particular, they aimed to identify the brain regions that are important for successful execution of expressive language in typically developing adolescents.
To this end, Authors first identified the multifrequency hubs by means of a modified version of the multilayer PageRank centrality and then reranked them according to their importance in fostering interlayer communication. 
Compared to standard single-layer analysis, this two-step procedure allowed to capture nonlinear interactions and resolve the task-related brain areas with a higher spatial resolution.
These regions mostly lied in the left hemisphere and represented possible conduits for interfrequency communication between action and perception systems that are crucial for language expression \citep{PULVERMULLER2018}.

Planning and executing motor acts is accompanied by changes in brain activity and connectivity on very short time scales of the order of milliseconds \citep{PFURTSCHELLER1999,SVOBODA2018}. \citet{tang_small-world_2010} used an EEG temporal network approach to characterize such fast brain functional organization during a simple foot movement task.
Compared to network sequences with randomly shuffled layers, brain networks showed a higher temporal clustering and a similar characteristic temporal path length (cf.~Sec.~\ref{sub:tool}).
Put differently, dynamic brain networks exhibited a temporal small-world propensity, supporting both segregation and integration of information through time.
While single-layer analysis had previously unveiled that segregation/integration properties fluctuate and adapt over the different phases of the movement \citep{de_vico_fallani_cortical_2008}, these findings provided new evidence on the intrinsic global temporal properties of motor-related brain networks.

\section{Multilayer network-based biomarkers of brain diseases}
\label{sec:brain_disorder}
Like any other complex system the brain can exhibit anomalous connectivity, which in turn may lead to abnormal behavior and clinical symptoms. 
Those brain connectivity changes can be spatially distributed, such as in schizophrenia or Alzheimer's disease, or localized such as in stroke or traumatic injuries \citep{HALLETT2020}. 
Looking at the network organization in both healthy and diseased conditions appears therefore fundamental to understand the resilience and vulnerabilities of the brain \citep{russo_neurobiology_2012}.
From a medicine perspective, network-based biomarkers would represent advanced tools to monitor the disease progression and inform new therapeutics to mitigate or counteract the effects of the disease.
In the last decade, standard network analysis has accumulated evidence documenting general reorganizational properties such as departure from optimal small-world configurations, aberrant modular reorganization, as well as significant loss of node centrality \citep{stam_modern_2014}.
Thus far these network changes have remained associated with a particular aspect, or layer, of information.
Since brain pathologies typically result from multifactor processes at different scales and levels, multilayer brain networks naturally constitute a more appropriate integrative modeling approach.
In the following, we present some of the most recent results obtained for different brain diseases, that provide new perspectives on the impacted multiscale network properties and can be used to improve diagnosis and prediction.


\subsection{Alzheimer's disease}

\begin{figure*}

\includegraphics[scale=0.6]{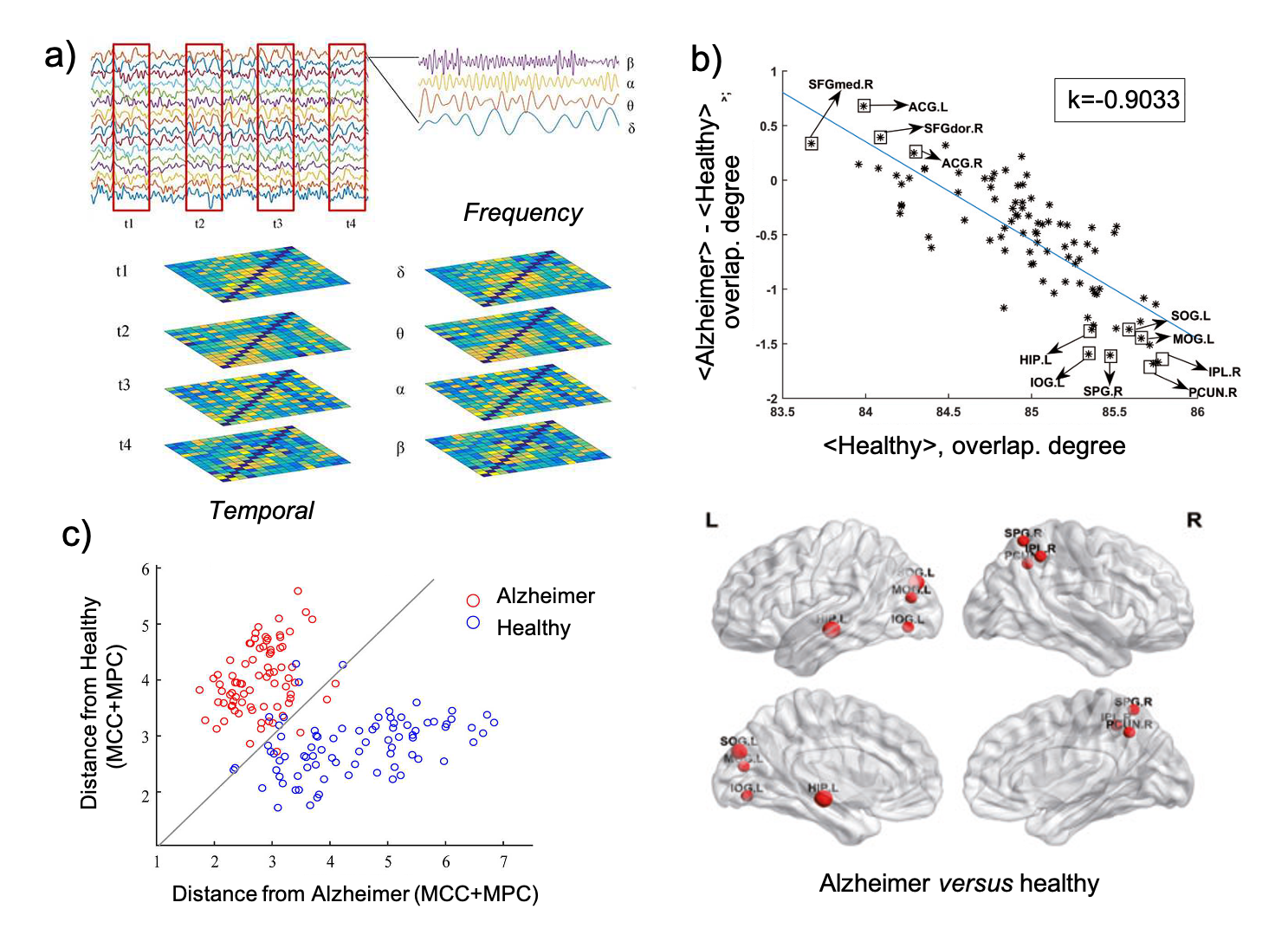}
\caption{\label{fig:AD_brain_organization} \textbf{Multifrequency and temporal reorganization of brain networks in Alzheimer’s disease.} Panel a) Multiplex brain networks are constructed by layering different frequency-specific networks, while temporal networks were constructed by concatenating time-specific networks within frequency bands. 
Panel b) Top: hub disruption of MEG multifrequency networks in patients with Alzheimer’s disease. Each point corresponds to a different brain area; k = slope of the regressing line. Bottom: Brain regions with significant between-group difference in overlapping weigthed degree. PCUN.R = right precuneus; HIP.L = left hippocampus; IPL.R = right inferior parietal, but supramarginal and angular gyri; SPG.R = right superior parietal gyrus; MOG.L = left middle occipital gyrus; SOG.L = left superior occipital gyrus; IOG.L = left inferior occipital gyrus.
Panel c) Scatter plot showing the Mahalanobis distance of each subject from AD or control group when combining of multiplex clustering coefficient (MCC) and participation coefficient (MPC) extracted from time-varying networks (gray line indicates equal distance).
Pictures and captions adapted from \citet{cai_functional_2020} (panel a, b) and from \citet{yu_selective_2017} (panel b), by permission of Oxford University Press for the latter.}

\end{figure*}

Alzheimer's disease (AD) is a neurodegenerative disorder and the most common form of dementia. Clinically, it is characterized by mild memory impairments that gradually evolve up to severe cognitive impairments and eventually to death. In $2016$, people affected by AD and other dementias were around $44$ million worldwide and this incidence is likely to augment because of longer life expectancy \citep{nichols_global_2019}.
At cellular level, AD is characterized by the progressive accumulation of $\tau$-tangles and $\beta$-amyloid plaques that cause neurons and synapses to die, thus leading to brain atrophy and disordered dysconnection patterns. 

While the consequences of these changes on large-scale brain networks have been widely investigated, the accumulated results are often discordant and depend on the considered spatial or temporal scale  \citep{TIJMS2013,gaubert_eeg_2019}. Multilayer networks represent an interesting approach to get an integrated, potentially more informative picture of the disease. 

Multiplex networks have been used to provide a unified description of AD brain reorganization across multiple  MEG frequency bands (\textbf{Fig.~\ref{fig:AD_brain_organization}a}).
\citet{yu_selective_2017} used different multiplex nodal metrics (e.g., overlapping clustering, local efficiency and betweenness centrality, cf.~Sec.~\ref{sub:tool}) and consistently showed that physiological multilayer hub regions, including posterior parts of the DMN, were severely impacted by AD (\textbf{Fig.~\ref{fig:AD_brain_organization}b}). 
Of note, these losses of functional hubs could not be observed when looking at individual frequency layers.
Such multilayer hub disruptions correlated with the accumulation of $\beta$-amyloid plaques in the cerebrospinal fluid, but also with the cognitive impairment of patients, demonstrating a potential clinical relevance.
By using the multiplex participation coefficient (Eq.~\ref{mplex_participation}), results indicated that most vulnerable hub regions in patients with AD also lost their ability to foster communication across frequencies compared to healthy control subjects.
Similar results were obtained independently by \citet{guillon_loss_2017}, showing a significant loss of multifrequency hubs in DMN regions and a strong association with memory impairment.
By using a classification analysis, they eventually showed that integrating multiparticipation coefficient values with equivalent single-layer network metrics leads to improved distinguishability of AD and healthy subjects. More recently, \citet{echegoyen_single_2021} showed that AD patients could be predicted by the lower values of algebraic connectivity $\lambda_2$ (cf.~Sec.~\ref{sub:multilayer_method}) in resting-state MEG multifrequency networks.
These results spot out new network mechanisms that hinder information load from flowing through different frequency bands and eventually impairs the cognitive abilities of AD patients.

\citet{cai_functional_2020}, addressed similar questions in EEG multifrequency brain networks. They showed that both multiplex clustering (Eq.~\ref{mplex_cluster}) and multiparticipation coefficients (Eq.~\ref{mplex_participation}) presented significant decrements with respect to healthy controls in the posterior areas of the brain.
These results confirmed a general tendency in AD patients to loose segregation and integration of information across signal frequencies. Yet, few observed increases in frontal areas suggested the presence of some compensatory mechanisms to be further elucidated \citep{guillon_disrupted_2019}.
In the same study, Authors also investigated the dynamic aspects of EEG brain networks in AD from a purely temporal perspective (\textbf{Fig.~\ref{fig:AD_brain_organization}a}). By using the aforementioned multilayer metrics, they showed that AD temporal segregation was mostly impacted by AD in frontal and occipital areas, while temporal integration properties were less affected as compared to healthy subjects, mainly because of its higher variability across nodes. However, when combined together, nodal values of temporal segregation and integration led to a very high discrimination between AD and healthy subjects ($>90\%$ accuracy), suggesting that spatial heterogeneity of temporal integration may also be related to progression of the disease (\textbf{Fig.~\ref{fig:AD_brain_organization}c}).

\begin{figure}
\includegraphics[scale=0.4]{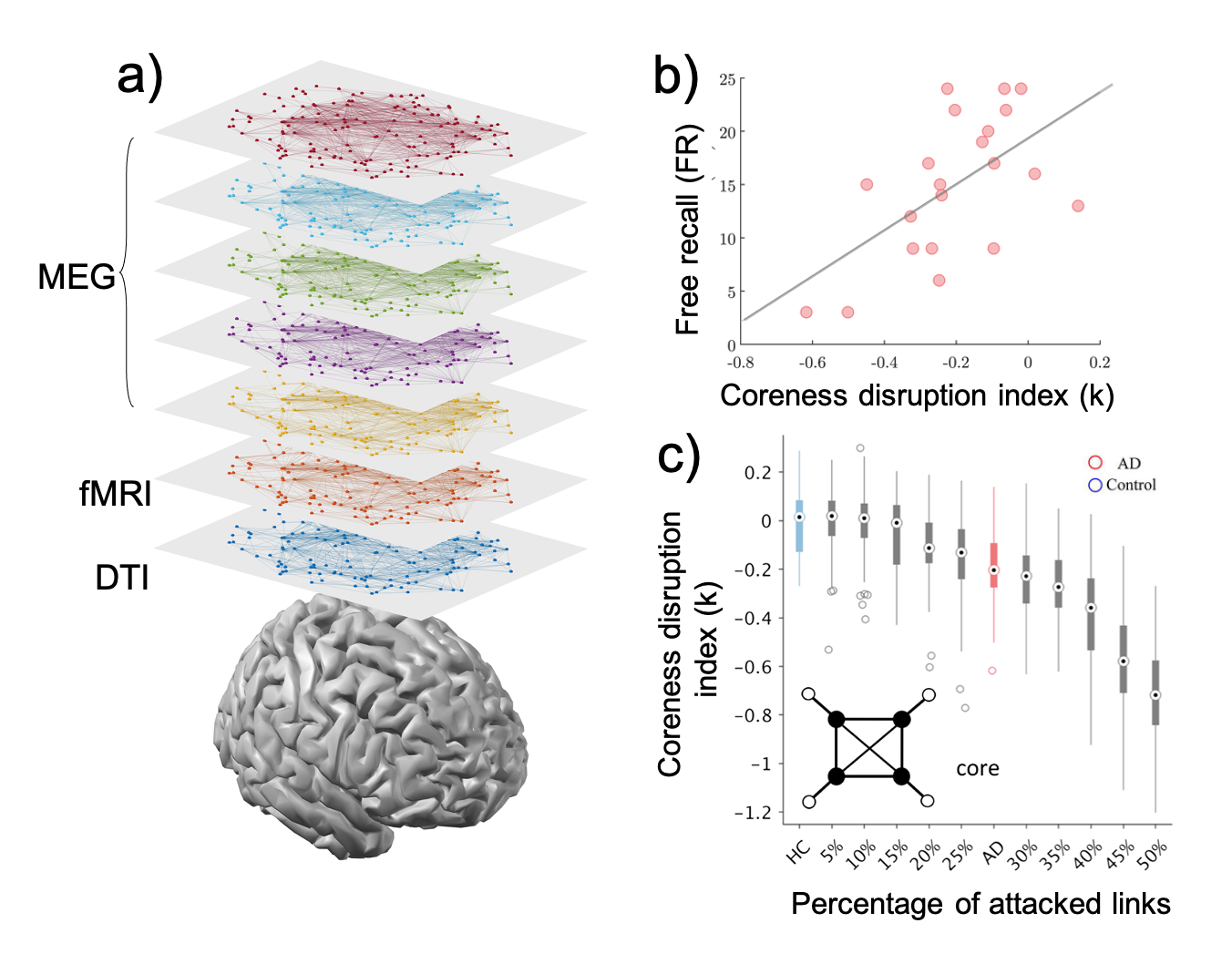}
\caption{\label{fig:AD_coreperiphery} \textbf{Multimodal brain networks reveal disrupted core-periphery structure in Alzheimer’s disease.} Panel a) Multimodal brain networks (multiplex) are constructed by layering DTI, fMRI, and several frequency-based MEG brain connectivity. 
Panel b) Spearman correlation (R = 0.59, p = 0.005) between the coreness disruption index ($\kappa$) and the memory impairment of AD patients as measured by the free recall (FR) test.
Panel c) Boxplots show the values of coreness disruption index ($\kappa$) obtained by progressively removing the edges preferentially connected to the multiplex periphery of the HC group. The blue (x-axis HC) and red (x-axis AD) boxplots illustrate respectively the $\kappa$ values for the HC and AD groups. 
Pictures and captions adapted from \citet{guillon_loss_2017} (Panel a) and \citet{guillon_disrupted_2019} (Panel b,c).}
\end{figure}

To integrate and disentagle the role of different neuroimaging modalities in AD, \citet{guillon_disrupted_2019} built multiplex networks composed of different connectivity types derived from DWI, fMRI and MEG data.
This represents so far the most complete type of multiplex brain network merging together structural and functional information (\textbf{Fig.~\ref{fig:AD_coreperiphery}a}).
By focusing on the mesoscale properties (cf.~Sec.~\ref{sub:tool}), Authors showed a selective reduction of multiplex coreness in the AD population, mainly involving temporal and parietal hub nodes of the DMN that are typically impacted by the anatomical atrophy and $\beta$-amyloid plaque deposition \citep{chetelat2010}. 
Such significant loss was mainly driven by few layers notably DWI, fMRI and MEG in the \textit{alpha1} ($7-10$ Hz) frequency range, and could be explained by a simple model reproducing the progressive random disconnection of the multilayer network via the preferential attacks of its core hubs (\textbf{Fig.~\ref{fig:AD_coreperiphery}b}).
From a clinical perspective, Authors eventually reported that patients with larger coreness disruption tended to have more severe memory and cognitive impairments, in line with the general tendency observed in other previously described studies \cite{yu_selective_2017} (\textbf{Fig.~\ref{fig:AD_coreperiphery}c}).
Recently, \citet{canalgarcia_multiplex_2022} built 2-layer multimodal networks from gray matter atrophy and amyloid deposition across different stages of AD in humans. Within a rigorous controlled study, they provided very specific results, not obtainable with traditional approaches from single imaging modalities. Notably, multiplex modularity (Eq.~\ref{eq:mplex_modularity}) revealed a characteristic module in the temporal brain area, likely to reflect the transition to AD dementia. Decreased values of multiplex participation coefficients (Eq.~\ref{mplex_participation}) in atrophy-related hub regions were also found in the later AD stage as compared to healthy controls. This study shed light on the non-trivial interplay between $\beta$-amyloid level and grey matter atrophy and its clinical relevance for AD.
 
Taken together, these results indicate that AD is characterized by a previously unappreciated multimodal and temporal dysconnection mechanism that primarly affects regions impacted by the atrophy process. Future research will be crucial to elucidate whether such a disruption tendency is compensated by other multilayer mechanisms, possibly involving more intact cortical systems, e.g.,  sensory motor \citep{guillon_disrupted_2019,kubicki2016,albers_at_2015}. 

\subsection{Neuropsychiatric disorders}

Among neuropsychiatric disorders, schizophrenia is certainly one of the most studied ones due to its large population incidence. In 2017, over 20 million people suffered from schizophrenia worlwide \citep{james_global_2018}. Typical clinical symptoms include hallucinations, emotional blunting and disorganized speech and thoughts. The biological causes of schizophrenia are still poorly understood and many hypotheses are currently being investigated based for example on neurotransmitter dysregulation \citep{lang_molecular_2007}, myelin reduction \citep{cassoli_disturbed_2015} as well as oxidative stress \citep{steullet_redox_2016}.
At large spatial scales, low and high frequency neuronal oscillations, as well as their interactions, have been widely documented as a core feature of the neuropathology underlying schizophrenia \citep{moran2011}.
Functional connectivity changes, within and between frequency bands, have been reported in schizophrenic patients \citep{sieben2013} and associated with persistent symptoms leading to disorganization of visuomotor mental functions \citep{brookes_multi-layer_2016}.

\begin{figure}
\includegraphics[scale=0.5]{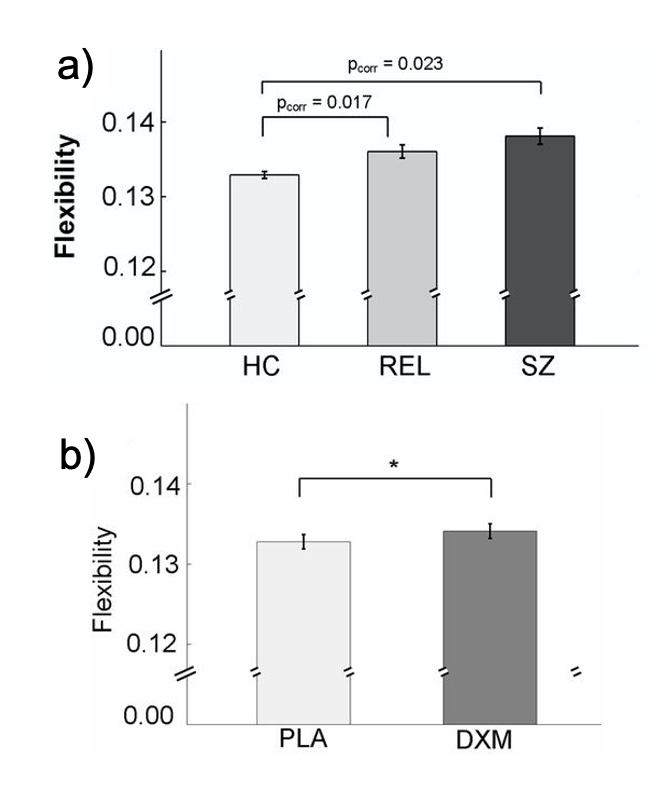}
\caption{\label{fig:schizophrenia} \textbf{Temporal network flexibility as a clinical marker of shizophrenia genetic risk.} Panel a) Significant increases in the mean dynamic reconfiguration of modular fMRI brain networks in unaffected first-grade relatives (gray bar, REL) and patients with schizophrenia (black bar, SZ) in comparison to matched healthy controls (white bar, HC) [F(2,196) = 6.541, P = 0.002]. Bars indicate mean values, and whiskers represent standard error means (SEMs).
Panel b) Significant increases in the mean dynamic reconfiguration of modular brain networks in healthy controls after application of dextrometorphan (DXM) [dark gray bars; repeated measures ANOVA placebo (PLA) versus DXM: F(1,34) = 5.291, P = 0.028] relative to PLA (light gray bars). 
Pictures and captions adapted from \citet{braun_dynamic_2016}.}
\end{figure}

By using a multiplex approach, \citet{domenico_mapping_2016} provided a first integrated characterization of the topological changes in schizophrenia from resting state fMRI-derived multifrequency networks.
In particular, they evaluated the multiplex PageRank centrality (cf.~Sec.~\ref{sub:tool}) and showed a substantial reorganization of the most important multifrequency hubs of the brain, such as the precuneus cortex, a key region for the basic physiological brain organization \citep{van_den_heuvel_network_2013}.
When injected into a random forest classifier, multiplex PageRank centrality metrics led to a classification accuracy of $80\%$, which is higher than standard network approaches, but comparable with otherwise much more sophisticated machine learning techniques.
At cellular levels, schizophrenia has been hypothesized to result from excitatory-inhibitory neuronal dysfunction, with a consequent abnormal temporal coordination between large-scale macro areas of the cerebral cortex \cite{uhl2013,uhlhaas_abnormal_2010}. By investigating temporal fMRI networks, \citet{braun_dynamic_2016}, showed that schizophrenic patients exhibited a multiplex network flexibility increase (cf.~Sec.~\ref{sub:tool}) with respect to healthy subjects during a working memory task, typically used to assess the neural basis of cognitive deficits \cite{meyer-lindenberg_regionally_2005,meyer2001} (\textbf{Fig.~\ref{fig:schizophrenia}a}).
Interestingly, Authors were able to reproduce the same \textit{hyperflexibility} when experimentally blocking the glutamate sensible synaptic receptors (NMDA receptors) in a separate group of healthy subjects (\textbf{Fig.~\ref{fig:schizophrenia}b}).
These results were further confirmed in a subsequent work, which localized such network hyperflexibility in specific brain zones including cerebellum, thalamus and frontoparietal task-related areas \citep{gifford_resting_2020}.
Altogether these findings indicated for the first time that microscale excitatory-inhibitory imbalances in schizophrenia might actually translate into temporally less stable and possibly disintegrated (rather than overly rigid) large-scale brain reorganization.

From a pure classification perspective multilayer brain networks have been also used as alternative multidimensional features to better discriminate between schizophrenic and healthy subjects.
\citet{lombardi_modelling_2019} considered a working memory fMRI experiment and built a 17-layers multiplex brain network where each layer contained a different type of nonlinear functional connectivity.
For each layer they extracted standard nodal centrality metrics (i.e., strength, betweenness, clustering, and PageRank) and used them as classification features.
Compared to a single-layer networks, built from simple linear correlations, they achieved a significantly higher classification ($\approx 90\%$ vs. $\approx 70\%$) for different types of working memory tasks.
Following the same goal, \citet{wilson_analysis_2020-2} considered resting state fMRI data over a group of healthy individuals and a group of patients with schizophrenia. 
Originally, they built for the two groups a multiplex brain network, where each layer represents the functional network of a specific individual.
By extending the popular \textit{node2vec} unsupervised network embedding procedure \citep{grover_node2vec_2016}, they learned continuous node feature representations from multilayer networks based on random walkers which are allowed to move across layers.
The resulting embeddings revealed a higher variability for the similarity between the nodes in the default mode network and salience subnetwork, suggesting a less stable within-module brain organization in the schizophrenic group.
While the overall classification accuracy did not outperform state-of-the-art performance, learning the features in an unsupervised approach might be nevertheless important for future applications in automatic diagnosis.

Major depressive disorder (MDD) is clinically characterized by severe fatigue, aphasia, difficulty to focus and suicidal thoughts in extreme cases. Symptoms are diverse and their severity largely differs among patients. Since effective treatments are currently available, scientific research mostly focuses on identifying predictive biomarkers to enable a more personalized therapeutics.
Previous studies suggested that MDD leads to several brain signal alterations affecting functional connectivity within but also between different frequency bands \cite{tian2019,nugent_multilayer_2020}.
To fully exploit this multifrequency information, \citet{dang_multilayer_2020} proposed a full multilayer approach to improve the diagnosis of MDD.
Specifically, they developed a convolutional neural network that directly takes as input the full multilayer brain networks to learn and extract the most discriminant features.  The resulting classification accuracy ($\approx 97\%$) was comparable to state-of-the-art methods based on specific frequency bands. While promising, these findings suggested that machine learning algorithms for multilayer brain networks still remain to be finetuned in view of their concrete implication in the identification of the best intervention strategy to cure or alleviate MDD-related symptoms. 

\subsection{Other neurological diseases}

Epilepsy is a group of neurological disorders characterized by seizures, which may vary in time and intensity, from short mild awareness loss to long vigorous convulsions. 
Epileptic seizures are underlied by excessive synchronized neuronal activity in the entire cerebral cortex or in parts of it.
In 2017, about $27$ million people suffered from epilepsy \citep{james_global_2018} among which $30\%$ are not curable with drug treatments \cite{kwan_early_2000}. 
Clinical research mostly aims at identifying predictive neural markers of the seizures to allow preventive treatments or to localize the origin of the seizure to inform precise surgery \citep{engel13}. 

Recent evidence has showed that epilepsy seizures are characterized by brain functional connectivity changes within, but also between, different brain signal frequencies  \citep{SAMIEE2018,villa2010,Jacobs2018}. From a topological perspective, decrements of network efficiency have been reported between low-high frequency bands, before the seizure onset, and were associated to sensorial disturbance and mild loss of consciousness \citep{yu_variation_2020}.

The intrinsic relationship between structural and functional layers can also unveil hidden connectivity structures characterizing different types of epilepsy. In this direction, \citet{huang_coherent_2020} used a DTI-fMRI multiplex approach to classify between epileptic seizures originating in different zones of the brain, namely the frontal and temporal lobe. 
In particular, Authors extended the concept of multiplex motifs to include subgraphs with more than $3$ nodes (cf.~Sec.~\ref{sub:tool}).
The most frequent multiplex patterns consisted of edges from both structural and functional layers that were spatially localized.
Notably, the structural components were quite stable across conditions and involved regions belonging to the DMN system (i.e., cuneus, precuneus, and peripheral cortex) \cite{horn2014}. 
Instead, the functional counterparts of the multiplex patterns, were highly variable and mostly involved regions  concentrated in the respective epileptogenic zones, i.e. temporal and frontal lobes. 
Eventually, Authors demonstrated the superiority of these multiplex connectivity patterns to discriminate between epileptic patients and healthy controls ($72$-$82\%$ classification accuracy), as compared to equivalent single-layer metrics or different multiplex metrics such as multiplex PageRank or algebraic connectivity (cf.~Sec.~\ref{sub:tool}).
These results are in line with the one-to-many relationships between structural and functional brain networks \citep{friston2013}, and can be used to finetune the research of predictive biomarkers in epilepsy. 

Consciousness disorders regroup a variety of symptoms which go from complete loss of awareness and wakefulness, such as coma, to minimal or inconsistent awareness \cite{giacino2014}. The differential diagnosis between the different types of disorders of consciousness is paramount to identify the best medical therapeutics.
Recent results suggest that frequency-dependent functional brain connectivity is crucial to characterize impairments of consciousness, as well as to predict possible recovery processes \cite{corazzol2017,chennu2014, cacciola2019}.
In an effort to provide an unified picture on the role of brain connectivity within and between frequency bands, \citet{naro_multiplex_2020}, adopted a multilayer network approach. By investigating brain networks derived from source-reconstructed EEG signals, Authors aimed to distinguish between patients suffering from unresponsive wakefulness syndrome (UWS) and minimally conscious state (MCS), which often present similar symptoms \citep{stender_diagnostic_2014}.
Results showed that several nodal multiplex metrics, including overlapping clustering, betweenness and multiplex participation coefficient (cf.~Sec.~\ref{sub:tool}), were significantly lower in UWS as compared to MCS patients. This was particularly evident in the frontoparietal regions of the brain whose relative loss of multiplex centrality is associated with the behavioral responsiveness of the patients quantified by the coma recovery scale \citep{gia2004}.
By adopting a full multilayer network approach, Authors eventually reported a significantly lower interlayer connection intensity in the UWS group and could spot out those patients who regained consciousness one year after the experiment. Notably, the discrimination between UWS and MCS patients was not observed when looking separately at frequency-specific network layers.
Although very preliminary, these results demonstrated the clinical value of considering multiplex/multilayer network approaches to derive more reliable neuromarkers of consciousness disorders.

\section{Emerging perspectives}

In the previous sections, we presented novel conceptual insights, tools and results that provide fresh perspectives on the intra/inter scale network properties of brain systems.
Research in the field is very active and many issues remain to be addressed in the future for ultimately characterizing multiscale/level brain organization. 
We then close this Colloquium by briefly focusing on three broad directions of advances in multilayer network theory that we believe particularly relevant for addressing this gap.

\subsection{Generative models of multiscale networks}
Generative models for brain networks allow to move from descriptive top-down approaches to mechanistic bottom-up ones \cite{betzel_generative_2017}. 
These models usually define a set of local connection rules (e.g., probabilistic rewiring or preferential attachment), to grow synthetic networks with specific global properties (e.g., small-worldness or scale-free degree distribution). 
Network models in neuroscience have been mostly driven by biological and topological evidences or hypotheses \citep{vertes_simple_2012,betzel_generative_2016}.

Biologically inspired models mainly implemented minimal wiring cost principles \citep{bullmore_economy_2012} and have been used to reproduce rich-club organization of brain networks \citep{vertes_generative_2014}, characterize the phase transition of axonal growth \citep{nicosia_phase_2013}, as well as to determine genetic risk factors of schizophrenia \citep{zhang_generative_2021}.
Topologically inspired models focused instead on reproducing the organizational properties of brain networks and have been adopted to identify the local connection mechanisms of network integration and segregation \citep{SIMPSON2012, obando_model_2017, SINKE2016}, or to reproduce the mesoscale modular properties of brain networks \citep{betzel_diversity_2018}.

The development of multilayer network models appears therefore a crucial step towards the multiscale modeling of the brain from a network perspective. On one hand, experimental technology is increasingly providing fresh data on different levels of neuronal interactions through 3D neuronal cultures \cite{hopkins2015}, calcium dynamics \citep{ahrens_whole-brain_2013}, spiking activity \citep{jun_fully_2017} and vascular support \citep{kirst_mapping_2020, mace_functional_2011} and might offer precious spatiotemporal insights to test biologically-plausible multilayer connection criteria.
On the other hand, we are currently witnessing a research thrust in the mathematical formalization of generative multilayer network models, mostly inspired by topological criteria.

For example, \citet{bazzi_framework_2019} recently proposed a unifying probabilistic framework to generate multiplex networks with any type of modular structure, that explicitly incorporates a user-specified tunable dependency between layers. These models might be useful to better quantify and understand the generation of mesoscale properties in multimodal and temporal brain networks.
Based on the extension of stochastic block models \citep{peixoto2014}, where nodes connect to each other with probabilities that depend on their group memberships, \citet{valles2016} proposed an original approach to derive the most probable multiplex modular network associated with any observed single layer network. Alternatively, \citet{lacasa_2018} provided a robust method relying on the Markovian diffusion of a random walker to determine whether a complex system is better modeled by a single interaction layer or by the interplay of multiple layers.
All these frameworks look particularly appealing for multiscale modeling as they might be used to identify the mesoscale inner workings of connectivity aggregation across different layers.
Finally, multilevel exponential random graph models potentially represent the most powerful framework due to their ability to characterize arbitrary connection patterns forming within and between layers, and to reproduce full multilayer networks \citep{wang2013}. This decade will be crucial to elucidate how multilevel biological knowledge and multilayer network tools can be merged to establish a new generation of network-based multiscale models of brain organization.

\subsection{Controllability of multilayer networks}
Understanding a complex system means being able to describe it, reproduce it and ultimately control it \cite{liu2016}.
In the last decade, the development of network control theory applied to brain connectivity has led to a paradigm shift, offering new tools to understand how the brain control itself and how it can be controlled by exogenous events \citep{tang_control_2018}.

Although still debated in the way it should be implemented and interpreted \cite{corbetta,jiang_irrelevance_2019}, network controllability has allowed to identify the \textit{driver} nodes that are more likely to steer the activity of human brain networks, opening huge possibilities for cognitive and clinical neuroscience, for example, via brain stimulation technology \cite{tang_control_2018, khambhati_virtual_2016, muldoon2016}.
More recently, a network control framework has been also used to determine the role of each \textit{C.~Elegans} neuron in locomotor behavior, that was confirmed by \textit{a-posteriori} laser ablations \citep{yan_network_2017}.
While the development of network controllability for single-layer systems is in its adolescence, its extension to multilevel systems is however still in its infancy.

The application to temporal networks is perhaps the most intuitive extension of structural controllability. By considering discrete time-varying linear dynamics of the system, \citet{posfai2014}, provided computational tools to study controllability based on temporal network characteristics. They specifically investigated the ability of single driver nodes to control a target and showed that the overall activity and the node degree distribution of the temporal network are the main features influencing controllability.
Although it might seem that static links would make it easier to control a system, \citet{li2017} demonstrated that temporal networks can be controlled more efficiently and require less energy than their static single-layer counterparts.
By using higher-order network models, \citet{zhang2021_control} also showed that the chronological ordering of interactions has a strong influence on the time needed to fully control the network. 

Determining the energy needed by the driver nodes to steer the system is also crucial. Excessively energetic control signals could be for example impossible to produce or could merely damage the system itself. In the case of full multilayer networks, \citet{wang_control_2017} demonstrated that there exists a tradeoff between the optimal controllability and optimal control energy that depends on the configuration and intensity of the interlayer connection patterns.
In a separate study, \citet{menichetti_control_2016} showed that controlling multiplex networks is more costly than controlling single layers taken in isolation, and that multiplex networks can exhibit stable controllability regardless of the stability of its layers. They also reported that multiplex networks need in general more drivers and that this number depends on the degree correlations between low-degree nodes in the different layers. Collectively, these findings encourage the development of controllability tools for multilayer brain networks with the goal of better disentangling the interaction between multiple scales, and improving the efficacy of possible intervention strategies.

\subsection{Machine learning and multilayer networks}

\begin{figure*}
\includegraphics[scale=0.5]{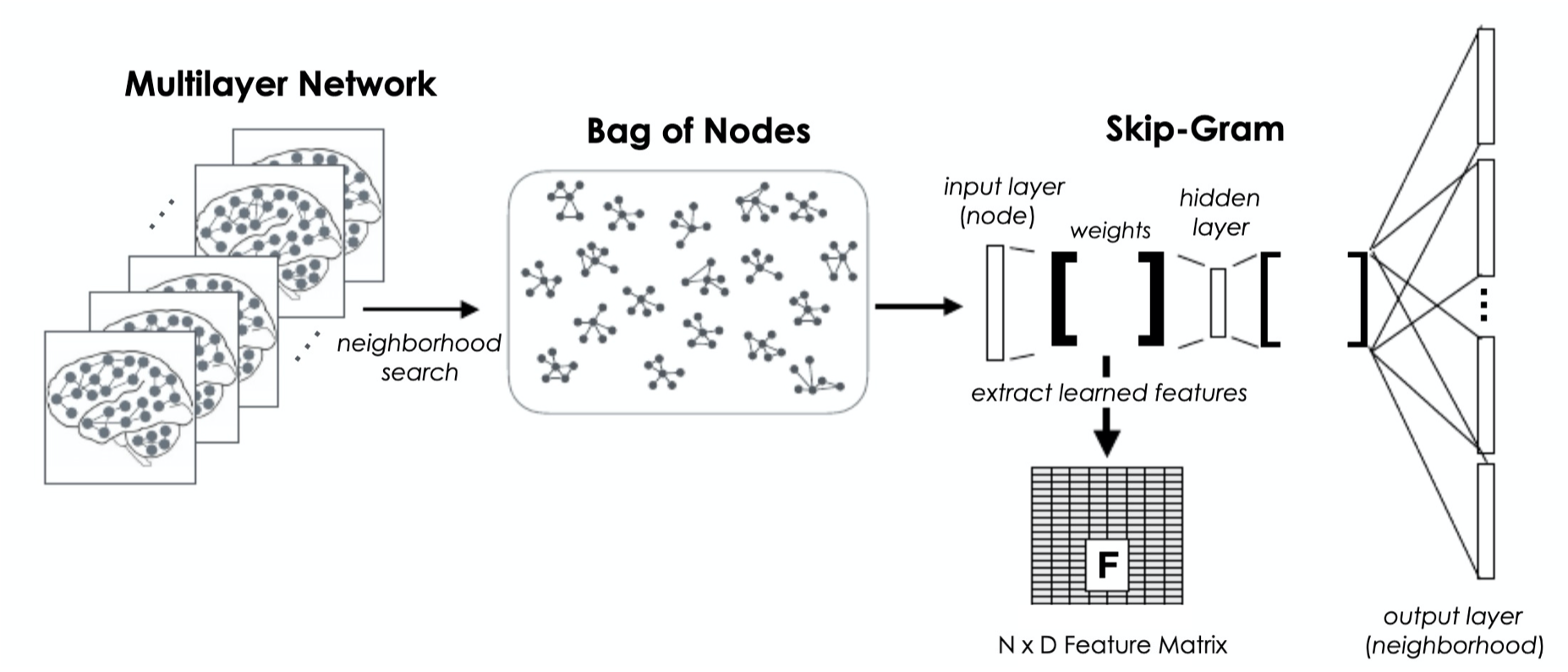}
\caption{\label{fig:node2vec} \textbf{Illustration of the multi-node2vec algorithm. 
} Beginning with a multilayer network (left), one first identifies a collection of multilayer neighborhoods (Bag of Nodes) via the NeighborhoodSearch
procedure. Next, the Optimization procedure calculates the maximum likelihood estimator F through the use of the Skip-Gram neural network model (right) on the identifed Bag of Nodes. Picture and caption adapted from \citet{wilson_analysis_2020-2}, reproduced with permission.}

\end{figure*}

Network science is a successful approach to analyze and model complex systems and uncover mechanisms that explain the emergence of functions. 
However, network theory alone often fails to efficiently manipulate large datasets as well as different levels of resolution. More importantly, it focuses on specific \textit{hand-crafted} topological features, and ignores less intuitive but possibly existing representative patterns, such as higher-order network interactions \cite{battiston2020}.

In this regard, machine learning represents a powerful technique to handle big amount of data and learn from the data itself the hidden patterns associated with the intrinsic phenomena of the system \citep{bishop2006}. 
As a counterpart, machine learning ignores the fundamental laws of physics and can result in ill-posed problems or non-interpretable solutions. 
The combination of machine learning and network science represents therefore a potential win-win strategy to address the above mentioned limitations, as recently demonstrated by a number of theoretical works and applications \citep{ZANIN2016, muscoloni_machine_2017}.
Nonetheless, when it comes to multiscale modeling, the type of algorithms must be rethought and extended to take into account the multilayer nature of the system, properly integrate the within- and between-layer concepts and explore the massive feature spaces \citep{domenico_modular_2015,alber_integrating_2019}.

Based on a specific class of deep learning algorithms, \citet{dang_multilayer_2020} developed a convolutional neural network that directly takes as input a full EEG multifrequency network to learn and extract the most discriminant features.
The core of their algorithm consisted of three consecutive convolutional layers, one batch normalization layer and one pooling layer. Such combination of basic hidden layers could effectively avoid overfitting and speed up the model training. Eventually, all learned features were concatenated together for classifying between healthy and major depressive diseased subjects. 

Machine learning can be optimized to operate feature engineering and embed the original multilayer network into a low-dimensional space so as to allow a minimal representation of the main intrinsic properties of the system.
Based on the popular \textit{node2vec} algorithm \citep{grover_node2vec_2016}, \citet{wilson_analysis_2020-2} introduced a fast and scalable extension, called \textit{multi-node2vec}, that learns the nodal features from complex multilayer networks through the Skip-Gram neural network model (\textbf{Fig.~\ref{fig:node2vec}}). This model was originally designed to extract the features of words' neighborhood in a text and was then adapted to characterize the neighborhood of nodes in a network \citep{mikolov_2013}. 
Applied to fMRI multisubject networks, Authors showed that it improves the visualization and clustering of brain regions into communities of similar features and discriminates between schizophrenic and healthy groups of subjects. 

More in general, the community detection task of partitioning the nodes of a multilayer network into densely connected subgroups, or communities, can be also viewed as a particular multilayer embedding. The development of multilayer community detection methods is still in its early stages, but several useful techniques have been developed over the past decade \citep{domenico_modular_2015,mucha2010,natalie2016,wilson2017}.






\section{Conclusion}
Understanding brain organization ultimately requires quantifying the interactions within and between multiple levels of neural structure and dynamics.
In the last decade, multilayer network theory has been introduced to characterize complex systems exhibiting different levels, or layers, of connectivity as well as cross-level interactions.
Here, we have presented and discussed many new developments in the field of multilayer network theory for the study of multiscale brain organization.
We anticipate that in conjunction with more accurate experimental technologies and increasing computational power, multilayer network theory can eventually become a key component of modern multiscale brain modeling.
Through this Colloquium, we hope to have provided fresh elements to stimulate new ideas in scientists and practioners wishing to advance multiscale brain modeling, which has profound implications for the bettering of our health and cognitive function.

\section{Acknowledgement}
We thank all the people with whom we have had formal and informal exchanges about this topic: A. Arenas, D. Bassett, F. Battiston, M. De Domenico, J. Gomez-Gardenes, J. Guillon, P. Hoevel, Y. Moreno, and P. Vertes, among others. Special thanks are given to A. Canal Garcia, M. Chavez, V. Latora, J. Martin-Buldu, and M. Serrano, who made useful suggestions on the initial version of this Colloquium. We also acknowledge Thibault Rolland for the graphical illustration preparation, and we are grateful to the anonymous referees whose reports allowed us to improve the previous manuscript. We acknowledge support from the European Research Council (ERC) under the European Union’s Horizon 2020 research and innovation program (Grant Agreement No. 864729). The funders had no role in the study design, data collection and analysis, decision to publish, or preparation of the manuscript.

\bibliography{colloquium_ref}
\end{document}